\ifx\onecol\undefined
    \documentclass[10pt,journal,twocolumn]{IEEEtran}
    \newcommand{\figsize}{0.90}
    \newcommand{\largeFigSize}{0.85}
\else 
    \documentclass[12pt,draftclsnofoot,journal,onecolumn]{IEEEtran}
    \newcommand{\figsize}{0.7}
\fi

\IEEEoverridecommandlockouts
\usepackage{amsmath}
\usepackage{amsthm,amssymb}
\usepackage{algorithm}
\usepackage{algorithmic}
\usepackage{bm,bbm}
\usepackage{balance}
\usepackage{braket}
\usepackage{booktabs}
\usepackage{color}
\usepackage{cases}
\usepackage{eqparbox}
\usepackage{enumitem}
\usepackage{graphicx}
\usepackage{mathrsfs}
\usepackage{multirow}
\usepackage{subfigure}
\usepackage{setspace}
\usepackage{threeparttable}
\usepackage{verbatim}

\usepackage[numbers,sort&compress]{natbib}

\newtheorem{theorem}{Theorem}
\newtheorem{remark}{Remark}

\usepackage{hyperref}
\hypersetup{hidelinks, 
colorlinks=true,
allcolors=black,
pdfstartview=Fit,
breaklinks=true}

\newcommand{\captionred}{} 
\newcommand{\red}[1]{{#1}}

\newcommand{\ri}{{\mathrm{i}}}
\newcommand{\diff}{{\mathrm{d}}}

\renewcommand{\Im}[1]{{\rm Im}{\{{#1}\}}}
\renewcommand{\Re}[1]{{\rm Re}{\{{#1}\}}}

\def \T {^{\mathsf{T}}}
\def \H {^{\mathsf{H}}}
\def \sinc {{\rm sinc}}

\begin{document}
\title{General Signal Model and Capacity Limit for Rydberg Quantum Information System }
\author{{Jieao~Zhu,~{\textit{Student Member, IEEE}}~and~Linglong~Dai,~{\textit{Fellow,~IEEE}}\vspace{-0.80cm} }
\thanks{This work was supported in part by the National Key Research and Development Program of China (Grant No. 2023YFB3811503), in part by the National Natural Science Foundation of China (Grant No. 62325106), and in part by the National Natural Science Foundation of China (Grant No. 62031019). }
\thanks{All authors are with the Department of Electronic Engineering, Tsinghua University, and the State Key Laboratory of Space Network and Communications, Tsinghua University, Beijing 100084, China (e-mails: zja21@mails.tsinghua.edu.cn, daill@tsinghua.edu.cn). }
\thanks{L. Dai is also with the EECS Department, Massachusetts Institute of Technology, Cambridge, MA 02139 USA. }
}

\maketitle

\begin{abstract}
    Rydberg atomic receivers represent a transformative approach to achieving high-sensitivity, broadband, and miniaturized radio frequency (RF) reception. 
    However, existing static signal models for Rydberg atomic receivers rely on the steady-state assumption of atomic quantum states, which cannot fully describe the signal reception process of dynamic signals. 
    To fill in this gap, in this paper, we present a general model to compute the dynamic signal response of Rydberg atomic receivers in closed form. 
    Specifically, by applying small-signal perturbation techniques to the quantum master equation, we derive closed-form Laplace domain transfer functions that characterize the receiver's dynamic responses to time-varying signal fields. 
    To gain more insights into the quantum-based RF-photocurrent conversion process, we further introduce the concept of quantum transconductance that describes the quantum system as an equivalent classical system. 
    By applying quantum transconductance, we quantify the influence of in-band blackbody radiation (BBR) noise on the atomic receiver sensitivity. 
    Extensive simulations for Rydberg atomic receivers validate the proposed signal model, and demonstrate the possibility of quantum receivers to outperform classical electronic receivers through the improvement of quantum transconductance. 
\end{abstract}

\section{Introduction} 

Classical electronic information systems have profoundly changed the world with their unprecedented capability of information acquisition and processing. Modern electronic systems, such as communication systems and radar systems, are capable of processing electromagnetic information, in which the physical process is mostly described by classical electrodynamics and thermodynamics. Unlike classical electronic systems, quantum information  systems provide the possibility of breaking the fundamental limits of classical electronic systems by leveraging quantum phenomena~\cite{nielsen2010quantum,gisin2007quantum}. 
For example, by exploiting the quantum superposition phenomena, quantum computing is expected to achieve an exponential acceleration in comparison to classical computing~\cite{nielsen2010quantum}. 
By utilizing the quantum no-cloning theorem, quantum key distribution can achieve unconditioned secure quantum communication~\cite{scarani2009security}. 
In addition, quantum sensing can achieve high sensitivity beyond the capability of classical sensors, e.g., quantum electrometers~\cite{degen2017quantum}. 

On the road to quantum communication and sensing, Rydberg atomic receiver is a promising technology that has recently gained much research attention~\cite{degen2017quantum,jing2020atomic,gong2024rydberg,cui2025towards,chen2025harnessing}. Unlike classical electronic receivers with metal antennas that convert radio-frequency (RF) signals to electronic signals, quantum receivers with Rydberg atoms convert RF signals to the change of atomic quantum states. 
Upon interacting with the external RF fields, the internal quantum states of these atoms are altered accordingly, resulting in a detectable change in the atom's optical transmission spectrum. This quantum phenomenon is known as the Autler-Townes (AT) effect~\cite{gray1978autler} which can be observed in an electromagnetically induced transparency (EIT) experiment~\cite{holloway2014sub}. The optical change is then captured by subsequent photodetectors to generate electronic signals that carry information about the incident RF fields. 



Due to the inherent high sensitivity of Rydberg states to RF fields, Rydberg atomic receivers can achieve weak signal detection that is expected to reach the quantum projection noise limit (QPNL) of $\sim 0.7\,{\rm nV\,cm^{-1}\,Hz^{-1/2}}$~\cite{jing2020atomic}, a level below the electronic thermal noise limit of around $3\sim 5\rm nV\,cm^{-1}\,Hz^{-1/2}$~\cite{fan2015atom}. In addition, unlike electronic receivers that are designed for specific bands, quantum receivers can be tuned from several MHz~\cite{yang2024pcb} to GHz~\cite{berweger2023rydberg} and THz~\cite{lin2023terahertz} with a single atomic vapor cell, which is impossible for state-of-the-art electronic receivers. Moreover, the small size of the atomic vapor cells (typically $\sim 2\,{\rm cm}$) enables wavelength-independent RF receiving that overcomes Chu's  limit~\cite{yuan2025electromagnetic} of classical antennas. These benefits of Rydberg atomic receivers~\cite{ZHANG20241515} make them a promising quantum information system for different applications in wireless communications~\cite{chen2025harnessing}, sensing~\cite{zhang2023quantum}, astronomy~\cite{graham2024rydberg}, radar~\cite{cui2025realizing}, etc.

\subsection{Prior works} 
The Autler-Townes effect was originally discovered by S. H. Autler and C. H. Townes in~\cite{autler1955stark}, where the response of atomic spectra to external resonant fields was discussed. The first Rydberg atom-based electrometer was attributed to~\cite{gordon2010quantum}, where an E-field measuring apparatus based on Rydberg atoms was proposed. As a general purpose E-field sensor, the Rydberg atomic receiver has been quickly considered in different applications~\cite{ZHANG20241515}, from metrology~\cite{sedlacek2012microwave} to wireless sensing~\cite{zhang2023quantum, schmidt24rydberg,cui2025realizing} and wireless communications~\cite{cai2023high,yuan2023rydberg,anderson2020atomic,berweger2023rydberg,holloway2020multiple}. 
Specifically, by leveraging the inherent sensitivity of Rydberg quantum sensors, the authors of~\cite{zhang2023quantum} improved the wireless sensing accuracy by an order of magnitude. 
For wireless communications, researchers have realized AM signal reception~\cite{yuan2023rydberg}, FM signal reception~\cite{anderson2020atomic},  QAM signal reception~\cite{holloway2019detecting}, multi-band reception~\cite{holloway2020multiple}, continuous-band reception~\cite{berweger2023rydberg}, and multi-carrier reception~\cite{cai2023high}. 
It should be pointed out that, most of these transmission techniques belong to the superheterodyne method~\cite{jing2020atomic,yuan2023rydberg,holloway2019detecting,yang2024pcb,holloway2020multiple}, which is the most representative method that can achieve the state-of-the-art atomic reception sensitivity~\cite{tu2024approaching} with both amplitude and phase detection capability. 

To understand the fundamental mechanisms and insights of atomic receivers, accurate signal models for each internal stage of the quantum system should first be established. 
The static signal model for atomic receivers was attributed to~\cite{jing2020atomic}, where the concept of intrinsic gain $\kappa$ (or intrinsic expansion coefficient~\cite{feng2023research}) was proposed to characterize the small-signal linear transfer coefficient from the signal Rabi frequency to the optical readout power. Based on this $\kappa$-coefficient, the wireless communication-oriented signal model for Rydberg atomic receivers was proposed in~\cite{gong2024rydberg}, where the atomic transfer relationship was derived by applying the static response~\cite{cui2025towards,cui2024multi} to external RF fields. Following this approach, the authors of~\cite{chen2025harnessing,wu2023linear} studied the large-signal case by treating the atomic receiver as a nonlinear RF envelope detector, and derived the linear range and channel capacities of the atomic receiver. Furthermore, to enable Rydberg atomic multiple-input multiple-output (MIMO) communications, the nonlinear point-to-point transmission model was extended to the MIMO transmission case~\cite{cui2025towards,cui2024mimo,cui2024multi}. 

However, the existing signal models for atomic receivers are static, i.e., they rely on the steady-state assumption~\cite{schmidt2024rydberg} that the input signal remains unchanged until the quantum system reaches its steady state. For practical signal transmission, dynamic signal models for Rydberg atomic receivers are required. This general model for dynamic signal reception is absent from the existing literature. We point out that the difficulty in establishing such a dynamic signal model is caused by the {\it nonlinear response} determined by the quantum master equation, i.e., the atomic optical response is generally not a linear function of its RF input.


\subsection{Our contributions} 
To fill in this gap, in this paper, we establish a general signal model that describes the dynamic Rydberg atomic response by applying the perturbation technique to linearize the master equation in the small-signal regime\footnote{Simulation codes will be provided to reproduce the results in this paper: \url{http://oa.ee.tsinghua.edu.cn/dailinglong/publications/publications.html}. }. The contributions of this paper are summarized as follows. 

\begin{itemize}
    \item Unlike traditional Rydberg analysis methods that rely on a static model, we propose a dynamic model for Rydberg atomic receivers by obtaining the perturbation solution to the quantum master equation. By analyzing this perturbation response with Laplace transform, a closed-form relationship between the optical response and the input RF signal is established, which is fully described by four Laplace domain transfer functions. \red{To ensure practicability of the model, the thermal Doppler effect is also incorporated in analytical form.} Based on these transfer functions, we propose the concept of {\it quantum transconductance} of a Rydberg atomic receiver, which is similar to the electronic transconductance of a transistor. 
    \item Based on the general model above, we analyze the atomic response to the inevitable in-band blackbody radiation (BBR) noise. Such analysis is not available for the existing static signal model. In addition to previous works that attribute noise to the imperfect optical readout and the standard quantum limit, we reveal that the BBR noise acts as another fundamental sensitivity performance limit to the atomic receivers, and that this sensitivity limit depends on a novel quantity called the BBR coherence factor. This coherence factor quantitatively describes the capability of BBR noise in corrupting the useful signal in the spatial domain. 
    \item Simulations are conducted for both single-input single-output (SISO) and MIMO Rydberg atomic communication systems, taking into consideration the general signal model with BBR noise and other noise sources. Extensive numerical results are presented to justify the proposed dynamic signal models in both the continuous-time and discrete-time domains. Numerical results have also demonstrated the possibility of constructing a coherent quantum-MIMO receiver with improved capacity compared to classical electronic receivers. 
\end{itemize}

\subsection{Organization and Notation}
\emph{Organization}:
The remainder of this paper is organized as follows. Section~\ref{sec2} introduces the basic quantum principles and system models of Rydberg atomic receivers. Section~\ref{sec3} computes the dynamic (transient) response of the atomic receivers to the time-varying external signal fields. Section~\ref{sec4} discusses the influence of BBR on the atomic response. Section~\ref{sec5} presents the equivalent baseband model and simulation results. Finally, conclusions are drawn in Section~\ref{sec6}.

\emph{Notation}: 
Bold uppercase characters ${\bf X}$ denote matrices, with $[{\bf X}]_{mn}$ representing its $(m,n)$--th entry; 
bold lowercase characters ${\bf x}$ denote vectors;
${\bf I}_n$ denotes the identity matrix of size $n$; 
${\bf X}\H, {\bf X}\T$, and ${\bf X}^*$ denotes Hermitian transpose, transpose, and complex conjugate of ${\bf X}$, respectively; 
for two operators $\mathcal{A}_1$ and $\mathcal{A}_2$, $[\mathcal{A}_1, \mathcal{A}_2]$ denotes the commutator $\mathcal{A}_1\mathcal{A}_2 - \mathcal{A}_2\mathcal{A}_1$, and $\{\mathcal{A}_1, \mathcal{A}_2\}$ denotes their anti-commutator $\mathcal{A}_1\mathcal{A}_2+\mathcal{A}_2\mathcal{A}_1$;  
$\mathcal{F}$ denotes Fourier transform, and $\mathcal{L}$ denotes the Laplace transform; 
For a time-domain signal $g(t)$, $g(\ri\omega)$ denotes its Fourier transform, and $g(s)$ denotes its Laplace transform; 
$\rm{i}$ is the imaginary unit; 
$\hbar$ denotes the reduced Planck's constant; 
$c_0$ denotes the speed of light in a vacuum; 
$\epsilon_0$ denotes the vacuum permittivity; 
$\eta_0$ denotes the vacuum wave impedance; 
${\rm vec}({\bf X})$ stacks the columns of the matrix $\bf X$ into a single column vector; 
${\rm PSD}[X(t)]$ denotes the double-sided power spectral density of the random process $X(t)$; 
For a real-valued bandpass signal $s(t)$, $s_a(t)$ denotes its analytic representation; 
$\mathcal{N}(\mu,\sigma^2)$ denotes the Gaussian distribution with mean $\mu$ and variance $\sigma^2$; 
$\mathcal{N}(\cdot|\mu,\sigma^2)$ denotes its probability density function.

\section{System Model}\label{sec2}
In this section, we first describe the signal processing pipeline of the Rydberg atomic receiver. Then, we describe the atomic response to external radio frequency (RF) fields. Finally, we discuss the end-to-end electro-optical DC/AC responses of the atomic receiver.

\subsection{Overall system description}
Consider a Rydberg atomic receiver with the RF-sensitive capability provided by an atomic vapor cell filled with alkali atoms, e.g., cesium-133 atoms. The atoms in the vapor cell are exposed to four electromagnetic waves: the probe light $E_{\rm p}(t)$ of angular frequency $\omega_p$, the control light $E_{\rm c}(t)$ of angular frequency $\omega_c$, the RF local oscillator (LO) signal $E_{\rm LO}(t)$ of angular frequency $\omega_{\rm LO}$, and the information-carrying RF signal $E_{\rm sig}(t)$. All the EM fields are assumed to be co-polarized and represented by its complex envelope. To achieve state-of-the-art E-field sensitivity performance, we adopt the superheterodyne architecture proposed in~\cite{jing2020atomic} throughout this paper. In this receiver architecture, the frequency of the signal field $E_{\rm sig}(t)$ is slightly different from that of $E_{\rm LO}(t)$, where the frequency difference is defined as the intermediate frequency $f_{\rm IF}$. 

\begin{figure*}
    \centering
    \includegraphics[width=\figsize\linewidth]{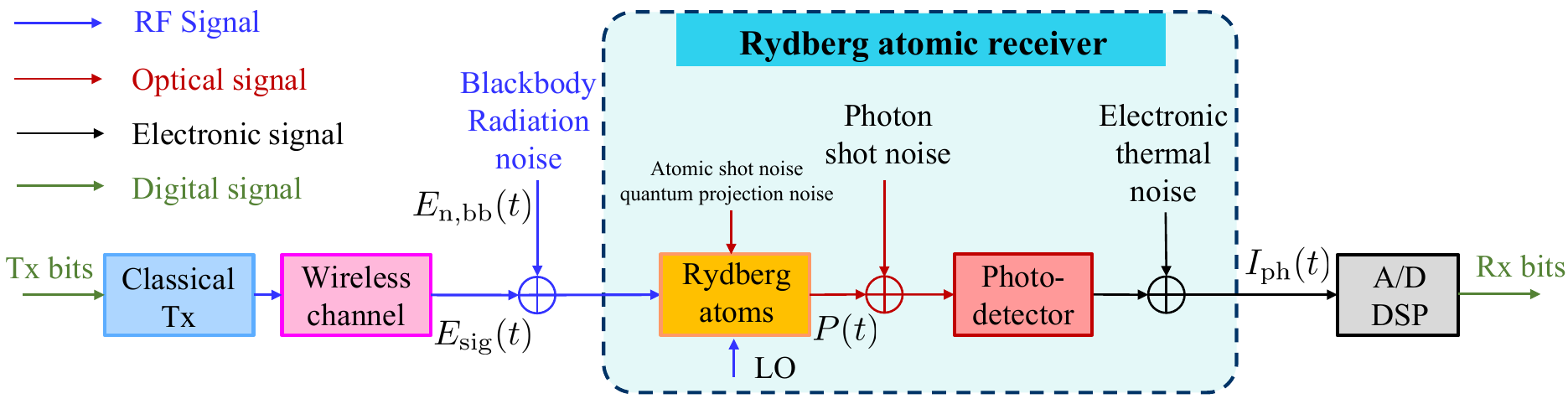}
    \caption{Signal reception pipeline of Rydberg atomic receivers. The Rydberg atoms act as an electro-optical mixer that down-converts the passband RF signal to the IF. The  Rydberg atomic receiver is composed of one atomic vapor cell and the subsequent analog-to-digital conversion (A/D) and digital signal processing (DSP) devices.} 
    \label{fig:Conceptual_RARE}
\end{figure*}

With a phase reference provided by the LO field, the time-varying signal field $E_{\rm sig}(t)$ interacts with the alkali atoms, resulting in a change in the transmission coefficient of the probe light. The transmitted probe light is then fed into a photodetector, e.g., a photodiode, to generate the IF electronic signal for further processing. The signal flow chart is shown in~Fig.~\ref{fig:Conceptual_RARE}, where it can be seen that the Rydberg atoms together with the photodetector act as a mixer~\cite{simons2019rydberg} that down-converts the passband RF signal to the intermediate frequency $f_{\rm IF}$ with the help of $E_{\rm LO}(t)$. 

\subsection{Quantum mechanisms for Rydberg atoms}
The probe/control lights and the RF signal mainly drive the electron transition between four quantum states, which is denoted as $\ket{1},\ket{2}, \ket{3},$ and $\ket{4}$. In this paper, the definitions of these energy levels are aligned to that of~\cite{jing2020atomic}, which are $6S_{1/2},F=4$; $6P_{3/2},F=5$; $47D_{5/2}$ and $48P_{3/2}$. 
This four-energy level quantum system is described by an $\hbar$-normalized Hamiltonian, which is given by 
\begin{equation}
{\bf H}(t)= \begin{bmatrix} 0 & \frac{\Omega_p}{2} & 0 & 0 \\ 
\frac{\Omega_p}{2} & -\Delta_p & \frac{\Omega_c}{2} & 0 \\ 
0 & \frac{\Omega_c}{2} & -\Delta_p-\Delta_c & \frac{\Omega_{\rm LO} + \Omega_{\rm sig}^*(t)}{2} \\ 
0 & 0 & \frac{\Omega_{\rm LO} + \Omega_{\rm sig}(t)}{2} & -\Delta_p-\Delta_c+\Delta_{\rm LO} \end{bmatrix},
\end{equation}
where $\Omega_p$, $\Omega_c$, and $\Omega_{\rm LO}$ are the Rabi frequencies of the probe field, control field, and the RF local oscillator (LO) field, respectively. $\Delta_p$, $\Delta_c$, and $\Delta_{\rm LO}$ are the frequency detunings from resonance, where a positive detuning means the frequency is higher than resonance. Due to the presence of intermediate frequency $f_{\rm IF}$, the signal Rabi frequency $\Omega_{\rm sig}(t)$ is a complex-valued function of time~\cite{jing2020atomic}, which is expressed as 
\begin{equation}
    \Omega_{\rm sig}(t)=\frac{\mu_{\rm RF}E_{\rm sig}(t)}{\hbar} = \frac{\mu_{\rm RF}E_{\rm sig, 0}}{\hbar} (I(t)+\ri Q(t)), 
    \label{eqn:sig_Rabi_freq}
\end{equation}
where $E_{\rm sig, 0}\,{\rm [V/m]}$ is some reference level of the incident electric field, and $\mu_{\rm  RF}\,{\rm [C\, m]}$ is the transition dipole between Rydberg states $\ket{3}$ and $\ket{4}$, and the real-valued oscillating RF field can be recovered from its analytic representation as 
\begin{equation}
    E_{\rm sig, real}(t) = E_{\rm sig,0}\cdot \Re{(I(t)+\ri Q(t))e^{\ri\omega_{\rm LO}t}}. 
\end{equation}

From a system-level viewpoint, the Rydberg atomic receiver operates as an electro-photonic mixer, where the beat frequency of the two RF signals $E_{\rm sig}(t)$ and $E_{\rm LO}(t)$ is amplified and modulated to the intensity variation of the transmissive probe light. The transmissive probe light intensity is expressed as 
\begin{equation}
    P(t) = P_{0}\exp{\left(-2 \int_{0}^{L} \alpha(x,t) {\rm d}x\right)}, 
    \label{eqn:probe_transmission}
\end{equation}
where $P_0\,[\rm W]$ is the incident probe power, $P(t)$ is the time-dependent transmissive probe power, $L\,{\rm [m]}$ is the atomic electro-optical interaction length that equals the length of atomic vapor cell, and $\alpha(x,t)\,{\rm [m^{-1}]}$ is the amplitude attenuation exponent per unit length at position $x\in [0,L]$ and time $t$. This attenuation exponent is related to the atomic quantum states via~\cite{jing2020atomic,sapiro2020time}
\begin{equation}
    \alpha(x,t)= - \frac{k_pN_0\mu_{12}^2}{\epsilon_0\hbar \Omega_p} \Im{\rho_{21}(x,t)},
    \label{eqn:attenuation_constant}
\end{equation}
where $k_p=\omega_p/c_0\,{\rm [m^{-1}]}$ is the wavenumber of the probe light, $N_0\,{\rm [m^{-3}]}$ is the atomic density inside the vapor cell, $\mu_{12}$ is the transition dipole from $\ket{1}$ to $\ket{2}$, and $\rho_{21}(x,t)$ is the $(2,1)$-th entry of the 4-by-4 Hermitian density matrix $\rho(x,t)$. \red{The temporal evolution of the density matrix for a single atom is fully described by the master equation 
\begin{equation}
    \frac{\partial \rho}{\partial t} = -\ri [{\bf H}, \rho] + D[\rho],  
    \label{eqn:master_equation}
\end{equation}
in which the decoherence operator $D[\rho]$ can be written as 
\begin{equation}
    D[\rho] = -\frac{1}{2}\{{\bf \Gamma}, \rho\} + {\bf \Lambda}[\rho], 
\end{equation}
where ${\bf \Gamma} = {\rm diag}([0, \gamma_2, \gamma_3, \gamma_4]+\gamma)$ describes the dephasing effects of the four-level system, ${\bf \Lambda}[\rho] = {\rm diag}(\gamma+\gamma_2\rho_{22}+\gamma_4\rho_{44}, \gamma_3\rho_{33}, 0, 0)$ is the re-population matrix, $\gamma$ is the transit dephasing rate~\cite{wu2023theoretical}, and $\gamma_i,i=2,3,4$ are the spontaneous decay rates of the $i$-th level~\cite{gong2024rydberg}. 
For Rydberg atoms in room temperature, the thermal movement of atoms causes Doppler effect of both the probe and control lasers, which further alters the density matrix $\rho$. The Doppler-aware density matrix is expressed as }
\red{
\begin{equation}
    \rho^{\rm D} = \int_{-\infty}^{\infty}\rho(\Delta_p-k_p v, \Delta_c+k_c v)\mathcal{N}(v|0, \sigma_v^2)\diff v,  \label{eqn:doppler_average_rho}
\end{equation}
}
\red{
where $v$ is the Doppler velocity parallel to the incident lasers, $k_p, k_c\,{\rm [m^{-1}]}$ are the wavenumbers of the probe and control lights, and $\sigma_v^2 = \sqrt{k_{\rm B}T/m_{\rm Cs}}$. Both numerical integration method~\cite{jing2020atomic} and analytical integration method~\cite{nagib2025fast} can be adopted to compute the Doppler ensemble average $\rho^{\rm D}$. 
}


As a first-order linear differential equation with variable coefficients, the master equation~\eqref{eqn:master_equation} determines the time evolution of $\rho_{21}(t)$. To better represent the time-domain signals, we decompose $\rho_{21}$ into two terms $\bar{\rho}_{21}+\Delta\rho_{21}(t)$, where $\bar{\rho}_{21}$ denotes the time-averaged value of $\rho_{21}(t)$, and $\Delta\rho_{21}(t)$ is the time-varying part. The probe transmission power $P(t)$ and the attenuation exponent $\alpha(x,t)$ admit similar decompositions as 
\begin{equation}
    \begin{aligned}
        P(t) &= \bar{P} + \Delta P(t), \\
        \alpha(x,t) &= \bar{\alpha}(x,t) + \Delta\alpha(x,t), 
    \end{aligned}
\end{equation}
where the time-averaged quantity $x$ is denoted by $\bar{x}$ in the rest part of this paper. After being detected by the photodiode, the output photocurrent $I_{\rm ph}(t)$ of the photodiode is given by 
\begin{equation}
    I_{\rm ph}(t) = \bar{I}_{\rm ph} + \Delta I_{\rm sig}(t) + \Delta I_{\rm n}(t), 
\end{equation}
where $\bar{I}_{\rm ph}$ is the average value of the output photocurrent, $\Delta I_{\rm sig}(t)$ is the photocurrent variation caused by the input E-field signal, and $\Delta I_{\rm n}(t)$ denotes the total noise current. The signal photocurrent is related to the optical signal power via 
\begin{equation}
    \Delta I_{\rm sig}(t) = \frac{q_e\eta}{\hbar\omega_p}\Delta P(t), 
\end{equation}
where $q_e$ is the elementary charge, $\eta$ is the quantum efficiency of the photodiode, and $\omega_p$ is the angular frequency of the probe light.

\subsection{Static DC and dynamic AC responses}
\begin{figure}[t]
    \centering
    \includegraphics[width=\figsize\linewidth]{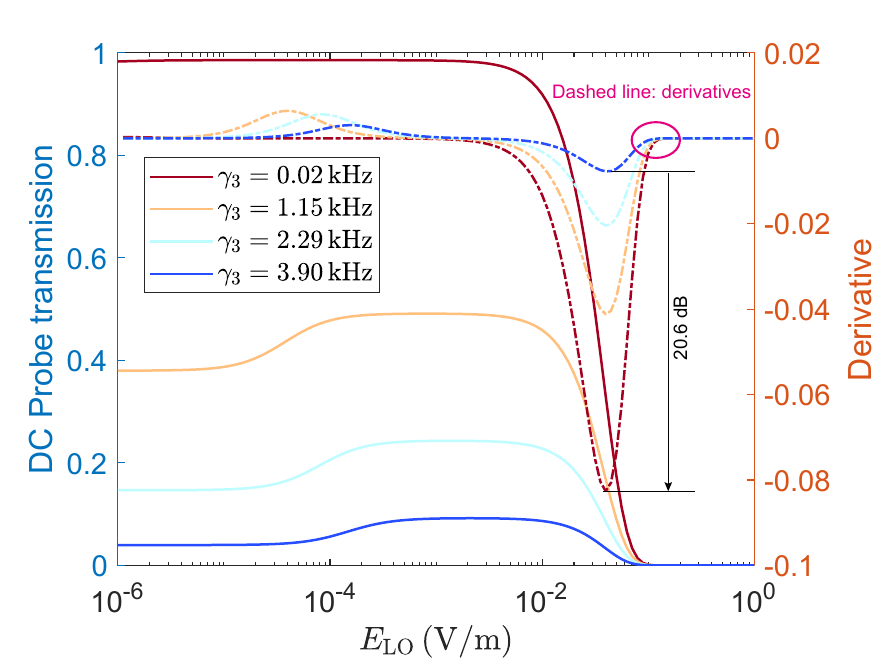}
    \caption{DC probe transmission $\bar{P}/P_0$ as a function of the LO E-field $E_{\rm LO}\,{\rm [V/m]}$ ($T=0\,{\rm K}$). Curves of the same color are respectively the $\bar{P}/P_0$-$E_{\rm LO}$ curve (solid) and its derivative (dashed). The value of $\gamma_4$ is proportionally adjusted with $\gamma_3$ during simulation. These curves show a strong dependence of the DC gain $\kappa$ on the inverse of Rydberg state lifetimes $\gamma_{3,4}$. }
    \label{fig:dc_gain}
\end{figure}

From the communication engineering viewpoint, we are particularly interested in the input-output relation of the Rydberg atomic system. This relation is canonically separated into two parts: the DC transfer gain in the form of a differential input-output slope, and the dynamic AC transfer function in the form of a frequency response $H(\ri\omega)$. The DC transfer gain $\kappa$ from the signal Rabi frequency $\Omega_{\rm sig}(t)$ to the probe power variation $\Delta P(t)$ is first defined by the authors of~\cite{jing2020atomic}, and has been studied theoretically in 2019 by the authors of~\cite{zhang2019detuning} and examined experimentally in~\cite{feng2023research}. Analytical expressions of the DC gain can be found in many papers, e.g., in~\cite{feng2023research, jing2020atomic, gong2024rydberg, zhang2019detuning,schmidt24rydberg,wu2024based}. In fact, we can apply Mathematica to symbolically solve the steady-state master equation $\diff \rho/\diff t=0$ for the steady-state solution $\bar{\rho}$ and extract its $(2,1)$-component to compute the DC probe transmission via~\eqref{eqn:attenuation_constant}. After obtaining the explicit expression of $\bar{\rho}_{21}$ as a function of $E_{\rm LO}$, the DC gain $\kappa$ can be computed by treating $E_{\rm sig}$ as small variations in $E_{\rm LO}$, which leads to 
\begin{equation}
    \kappa = \frac{\partial \bar{P}}{\partial E_{\rm LO}}\frac{\partial E_{\rm LO}}{\partial \Omega_{\rm LO}} = \frac{\hbar}{\mu_{\rm RF}} \frac{\partial \bar{P}}{\partial E_{\rm LO}}.   
\end{equation}
Thus, the DC gain is determined by the slope, i.e., the derivative, $\partial\bar{P}/\partial E_{\rm LO}$ of the probe transmission-LO strength curve, as shown in Fig.~\ref{fig:dc_gain}. 

Note that the dynamic AC response~\cite{bohaichuk2022origins} can behave very differently from its DC counterpart, although the dynamic response is expected to be in agreement with the static response at $\omega=0$. Since key communication performance indicators, like the system bandwidths and noise power spectral density (PSD), are fundamentally determined by the dynamic response of the system, we are devoted to evaluating the dynamic transfer function in the next section. \red{Detailed comparison between the static transfer gain $\kappa$ and the dynamic transfer function $\kappa(\ri\omega)$ is discussed in {\bf Section~\ref{sec3-5}. }
}

\section{Dynamic Models}\label{sec3}
In this section, we find that the atomic receiver can be treated, up to a first-order approximation, as a linear time-invariant (LTI) system. Specifically, we solve the transfer function of Rydberg atomic receivers by applying the small-signal perturbation technique to the master equation. The small-signal approximation holds when $E_{\rm sig}\ll E_{\rm LO}$.  

\subsection{Reformulation of the master equation}
Since the master equation~\eqref{eqn:master_equation} is a first-order linear ordinary differential equation (ODE) in variable $t$, it can be reformulated into the standard matrix-vector form as 
\begin{equation}
    \frac{{\rm d}{\bf x}_\epsilon}{\diff t} = ({\bf A}_0+\epsilon {\bf A}_1(t)){\bf x}_\epsilon(t),
    \label{eqn:reformulated_master_eqn}
\end{equation}
where ${\bf x}_\epsilon(t) = {\rm vec}(\rho)\in\mathbb{C}^{16}$ is the vectorized density matrix under small Hamiltonian perturbation $\epsilon{\bf A}_1(t)$, ${\bf A}_0$ is the unperturbed vectorized Hamiltonian given by 
\red{
\begin{equation}
\begin{aligned}
  {\bf A}_0 &= -\ri ({\bf I}_4\otimes {\bf H}_0- {\bf H}_0\T\otimes {\bf I}_4) - \frac{1}{2}({\bf \Gamma}\otimes {\bf I}_4 + {\bf I}_4\otimes {\bf \Gamma}) \\
  &+ \gamma{\bf E}_{1,1}+(\gamma+\gamma_2) {\bf E}_{1,6}+(\gamma+\gamma_4) {\bf E}_{1, 16} + \gamma_3 {\bf E}_{6, 11} \\ 
  &+ \gamma {\bf E}_{1,11}, 
\end{aligned}
\end{equation} }
${\bf H}_0$ is the unperturbed Hamiltonian without the $\Omega_{\rm sig}(t)$ term, and ${\bf E}_{ij}\in\mathbb{C}^{16\times 16}$ denotes the matrix with the only ``1'' entry located at $(i,j)$-th position. In the context of atomic receivers, the perturbation term is defined by $\Omega_{\rm sig}(t)$ via
\begin{equation}
    \begin{aligned}
        \epsilon {\bf A}_1(t) = -\frac{\ri}{2}&\left[  {\bf I}_4\otimes (\Omega_{\rm sig}(t){\bf E}_{43}+\Omega_{\rm sig}^*(t){\bf E}_{34}) \right.\\
        &\left.- (\Omega_{\rm sig}(t){\bf E}_{34}+\Omega_{\rm sig}^*(t){\bf E}_{43})\otimes{\bf I}_4\right]
    \end{aligned}
\end{equation}
Generally, the reformulated master equation~\eqref{eqn:reformulated_master_eqn} can be solved by matrix exponents in the unperturbed case $\epsilon=0$, where the unperturbed solution is 
\begin{equation}
    {\bf x}_0(t) = \exp(t{\bf A}_0){\bf x}(0).
\end{equation}
Note that the solution to ~\eqref{eqn:reformulated_master_eqn} may be singular, i.e., a finite-energy perturbation $\epsilon{\bf A}_1(t)$ may lead to an infinite-energy differential response ${\bf x}_\epsilon(t)-{\bf x}_0(t)$. This phenomenon is caused by the zero eigenvalue of ${\bf A}_0$, i.e., there exists some steady-state solution $\bar{\bf x} \in\mathbb{C}^{16}$ that satisfies ${\bf A}_0 \bar{\bf x}={\bf 0}$. Mathematically, this singularity can be eliminated by introducing an orthonormal basis transform ${\bf Q}$ to the state vector ${\bf x}_\epsilon$, which is expressed as 
\begin{equation}
    {\bf y}_\epsilon(t):= {\bf Q}\T {\bf x}_\epsilon(t).
\end{equation}
The orthonormal transform matrix ${\bf Q}$ is designed to be 
\begin{equation}
    {\bf Q} = [{\bf u}_4, {\bf q}_1, \cdots, {\bf q}_{15}],
\end{equation}
\red{where ${\bf u}_4\in\mathbb{R}^{16}$ with four ``1/2'' entries located in the 1, 6, 11, 16-th positions (or ${\bf u}_4={\rm vec}({\bf I}_4)$~\cite{wu2024based})}, and ${\bf q}_k, k=2,\cdots, 15$ are arbitrarily chosen unit vectors that makes ${\bf Q}$ an orthonormal matrix. The choice of ${\bf u}_4$ is related to the trace-preserving property of the master equation~\eqref{eqn:master_equation} that ensures ${\bf u}_4\T {\bf A}_0 = {\bf 0}$. The transformed master equation is expressed as 
\begin{equation}
\begin{aligned}
    \frac{\diff {\bf y}_\epsilon(t)}{\diff t} &= {\bf Q}\T({\bf A}_0+\epsilon {\bf A}_1(t)){\bf Q}{\bf y}_\epsilon(t) \\ 
    &:= ({\bf B}_0 + \epsilon{\bf B}_1(t)){\bf y}_\epsilon(t). 
\end{aligned}
\end{equation}
Due to the trace-preserving property of~\eqref{eqn:master_equation}, the structure of the matrix ${\bf B}_0$ is 
\begin{equation}
    {\bf B}_0 = \begin{bmatrix} 0 & {\bf 0}_{1\times 15} \\ {\bf w}_0 & {\bf C}_0 \end{bmatrix},
\end{equation}
where ${\bf w}_0\in\mathbb{C}^{15}$ and ${\bf C}_0\in\mathbb{C}^{15\times 15}$ are constant matrices determined by ${\bf H}_0$ and the choice of ${\bf Q}$. The same structure also applies to the perturbation term $\epsilon {\bf B}_1(t)$, yielding the definitions of $\epsilon{\bf w}_1(t)$ and $\epsilon{\bf C}_1(t)$. From the structure of ${\bf B}_0+\epsilon{\bf B}_1(t)$, it can be seen that $\diff [{\bf y}_\epsilon]_1/\diff t=0$, which yields $[{\bf y}_\epsilon]_1(t)\equiv [{\bf y}_\epsilon]_1(0) = {\bf u}_4\T{\bf x}(0)=1/2$. To study the time-varying properties of the remaining components of ${\bf y}_\epsilon$, we let ${\bf z}_\epsilon = [{\bf y}_{\epsilon}]_{ 2:16}\in\mathbb{C}^{15}$. This yields, for any $\epsilon>0$, 
\begin{equation}
    \frac{\diff {\bf z}_\epsilon}{\diff t} = \frac{1}{2}{\bf w}_0 + ({\bf C}_0+\epsilon{\bf C}_1(t)){\bf z}_\epsilon(t), \label{eqn:stable_master_equation}
\end{equation}
where the small-signal term $\epsilon{\bf w}_1(t)$ can be eliminated by properly choosing ${\bf q}_k$. One can check that all of the eigenvalues of ${\bf C}_0$ have negative real parts, indicating that~\eqref{eqn:stable_master_equation} is non-singular against small perturbations $\epsilon {\bf C}_1(t)$.

\subsection{Small-signal transfer function}
After reformulation of the master equation into the stable form~\eqref{eqn:stable_master_equation}, the perturbation method can be applied to evaluate the small-signal transfer function from the entries of $\epsilon{\bf C}_1(t)$ to the variations that appear in $\rho_{21}(t) = \left[{\rm unvec}({\bf Q}[1/2, {\bf z}(t)\T]\T)\right]_{21}$. Starting from~\eqref{eqn:stable_master_equation}, we assume the $\epsilon$-perturbed solution to be 
\begin{equation}
    {\bf z}_\epsilon(t) = {\bf z}_0(t)+\epsilon \frac{\partial {\bf z}_\epsilon}{\partial\epsilon}\bigg\vert_{\epsilon=0} + o(\epsilon).
\end{equation}
Substitute this first-order approximated solution into~\eqref{eqn:stable_master_equation}, we get
\begin{equation}
\begin{aligned}
    \frac{\diff {\bf z}_0}{\diff t} + \epsilon \frac{\partial}{\partial t}\frac{\partial {\bf z}_\epsilon}{\partial\epsilon} &= \frac{1}{2}{\bf w}_0+{\bf C}_0{\bf z}_0(t) \\
    &+\epsilon \left({\bf C}_1(t){\bf z}_0(t)+{\bf C}_0\frac{\partial {\bf z}_\epsilon}{\partial\epsilon}\right) + o(\epsilon),
\end{aligned}
\end{equation}
where $o(\epsilon)$ denotes terms that vanish after dividing by $\epsilon$ as $\epsilon\to 0$. We know that ${\bf z}_0(t)$ satisfies the unperturbed differential equation $\diff  {\bf z}_0/\diff t={\bf C}_0{\bf z}_0 + {\bf w}_0/2$, thus we get 
\begin{equation}
    \frac{\partial}{\partial t}\left(\frac{\partial {\bf z}_\epsilon}{\partial \epsilon}\right) = {\bf C}_1(t){\bf z}_0(t) + {\bf C}_0\left(\frac{\partial {\bf z}_\epsilon}{\partial\epsilon}\right). 
\end{equation}
This is a first-order constant-coefficient matrix-vector differential equation of unknown vector $\partial{\bf z}_\epsilon/\partial\epsilon$, where the solution can be expressed as 
\begin{equation} 
    \frac{\partial {\bf z}_\epsilon}{\partial \epsilon}\bigg\vert_{\epsilon=0}(t) = \int_0^t \exp \left((t-\tau){\bf C}_0\right){\bf C}_1(\tau){\bf z}_0(\tau)\diff\tau.
    \label{eqn:time_domain_response}
\end{equation}
The equation~\eqref{eqn:time_domain_response} describes the first-order dynamic response of the Rydberg atomic receiver, expressed in terms of a convolution of the matrix exponential $\exp(t{\bf C}_0)$ with the product of the input excitation term ${\bf C}_1(t)$ and the unperturbed solution ${\bf z}_0(t)$. Multiply both sides of~\eqref{eqn:time_domain_response} with $\epsilon$ and take the Laplace transform, the $s$-domain small-signal response of the atomic receiver is given by 
\begin{equation}
\begin{aligned}
    \mathcal{L}[\epsilon \frac{\partial {\bf z}_\epsilon}{\partial \epsilon}](s)  &\overset{(a)}{=} \mathcal{L}[e^{t{\bf C}_0}](s)\cdot (\epsilon{\bf C}_1(s)) \cdot \bar{\bf z}_0 \\ 
    &= (s{\bf I}_{15} - {\bf C}_0)^{-1}(\epsilon {\bf C}_1(s))\bar{\bf z}_0,
\end{aligned}
\end{equation}
where (a) comes from the assumption that the Rydberg atomic system have reached steady-state before $E_{\rm sig}$ is applied. The term $\epsilon{\bf C}_1(s)$ is the Laplace transform of the time-domain perturbation $\epsilon {\bf C}_1(t)$, and $\bar{\bf z}_0$ is the steady-state solution to~\eqref{eqn:stable_master_equation} with $\epsilon=0$, which is given by. 
\begin{equation}
    \bar{\bf z}_0 = -\frac{1}{2}{\bf C}_0^{-1}{\bf w}_0. 
\end{equation}
By undoing the applied linear transforms in ${\bf z}_\epsilon(t)$, the transfer function of any variation in the Hamiltonian $\epsilon{\bf H}_1(t)$ to the variations in any element of the density matrix $[\Delta\rho_{mn}]_{4\times 4}$ can be easily computed. In the case of Rydberg atomic receivers, we are particularly interested in $\Delta\rho_{21}$, since it fully describes the probe transmission coefficient. 

Let ${\bf T}(s) := [T_{k\ell}(s)]_{4\times 4}$ denote the transfer functions from $\epsilon{\bf H}_1(t)$ to $\Delta\rho_{21}$. Then, ${\bf T}(s)$ is expressed as 
\begin{equation}
    {T_{k\ell}}(s)=\left[{\bf Q}\begin{bmatrix} {\bf 0} \\ {\bf I}_{15}\end{bmatrix} (s{\bf I}_{15}-{\bf C}_0)^{-1}  \frac{\partial (\epsilon {\bf C}_1)}{\partial [\epsilon {\bf H}_1]_{k\ell}} \bar{\bf z}_0 \right]_{2}. 
    \label{eqn:Tkl_formula}
\end{equation}
Since the input signal only alters $[{\bf H}_1]_{34}$ and $[{\bf H}_1]_{43}$, the two entries $T_{34}$ and $T_{43}$ contribute to the end-to-end electro-optical response of the atomic receiver. Rewrite the signal Rabi frequency~\eqref{eqn:sig_Rabi_freq} in terms of the analytic baseband signal $X(t) := I(t) + \ri Q(t)$, then the Laplace transform of the atomic response $\Delta\rho_{21}(t)$ is expressed as 
\begin{equation}
\begin{aligned}
    \mathcal{L}[\Delta\rho_{21}](s) &= \frac{\mu_{\rm RF}E_{\rm sig,0}}{2\hbar} \\
    &\times \left[T_{34}(s)(I(s)-\ri Q(s))+T_{43}(s)(I(s)+\ri Q(s)) \right],
\end{aligned}\label{eqn:transfer_function_IQ_to_rho}
\end{equation}
where $I(s)$ and $Q(s)$ are Laplace transforms of $I(t)$ and $Q(t)$, respectively. We can separate the total gain into two parts: The in-phase gain $G_{\rm I}(s)\,{\rm [Hz^{-1}]}$ and the quadrature gain $G_{\rm Q}(s)\,{\rm [Hz^{-1}]}$, which are given by 
\begin{equation}
\begin{aligned}
    G_{\rm I}(s) &= \frac{1}{2}(T_{43}(s)+T_{34}(s)) \\
    G_{\rm Q}(s) &= \frac{\ri }{2}(T_{43}(s)-T_{34}(s)). 
    \label{eqn:GI_GQ}
\end{aligned}
\end{equation}
To separate the real-part response of $\Delta\rho_{21}$ from the imaginary-part response, we separate the real-part filter and imaginary-part filter of both $G_{\rm I}(s)$ and $G_{\rm Q}(s)$ as 
\begin{equation}
    \begin{aligned}
        G_{\rm I}(s) &= G_{{\rm I}, 1}(s) + \ri G_{{\rm I}, 2}(s), \\ 
        G_{\rm Q}(s) &= G_{{\rm Q}, 1}(s) + \ri G_{{\rm Q}, 2}(s).
    \end{aligned}
\end{equation}
This separation is done by extracting the real- and imaginary-part of the numerator polynomial coefficients of $G_{\rm I}(s)$ and $G_{\rm Q}(s)$. Note that the system characteristic polynomial (denominator polynomial) $p(s):=\det (s{\bf I}_{15}-{\bf C}_0)$ is always a real-coefficient polynomial, i.e., the complex system poles always appear in conjugate pairs. With these four separated gains, the real- (imaginary-) part response of $\Delta\rho_{21}$ is given by
\begin{equation}
    \mathcal{L}\begin{bmatrix}\Re{\Delta\rho_{21}} \\ \Im{\Delta\rho_{21}}\end{bmatrix} = \frac{\mu_{\rm RF}E_{\rm sig,0}}{\hbar}\begin{bmatrix}G_{\rm I,1}(s) & G_{\rm Q,1}(s) \\ G_{\rm I,2}(s) & G_{\rm Q,2}(s) \end{bmatrix} \cdot\begin{bmatrix} I(s)  \\ Q(s) \end{bmatrix} \label{eqn:re_im_transfer_functions}
\end{equation}
In the near-resonance regime where the laser detunings, the RF LO detuning and the intermediate frequency $f_{\rm IF}$ are all small, we numerically observe that $G_{\rm Q}$ is significantly smaller than $G_{\rm I}$, and $G_{\rm I}$ is nearly purely imaginary. This means the $G_{{\rm I},2}$ component is significantly larger than other three components, i.e., the Rydberg atoms respond only to the in-phase component of $E_{\rm sig}$ and ignore the quadrature component. This observation is somehow in accordance with the non-coherent amplitude receiving model that appears in~\cite{cui2025towards,gong2024rydberg}. 



\subsection{Incorporating the thermal Doppler effect}
\red{
For Rydberg atoms in room temperature, the EIT lines are broadened by several hundreds of MHz by thermal motion, which has a significant impact on the atomic response to external RF signals. Similar to the difference between $\rho$ and $\rho^{\rm D}$ for the static response, the transfer functions $T_{k\ell}(s)$ in~\eqref{eqn:Tkl_formula} for the dynamic response should also be updated, which is expressed as 
\begin{equation}
    T_{k\ell}^{\rm D}(s) := \int_{-\infty}^\infty T_{k\ell}(s,v)\mathcal{N}(v|0, \sigma_{v}^2) \diff v, \label{eqn:Tkl_with_Doppler}
\end{equation}
where, following~\eqref{eqn:Tkl_formula}, we have 
\begin{equation}
    T_{k\ell}(s,v) = \left[{\bf Q}\begin{bmatrix}{\bf 0} \\ {\bf I}_{15}\end{bmatrix} \left(s{\bf I}_{15}-{\bf C}_0 - \frac{v}{\sigma_v}{\bf C}_v\right)^{-1} {\bf F}_{k\ell}\bar{\bf z}_v\right]_2, \label{eqn:Tkl_s_v}
\end{equation}
${\bf F}_{k\ell}$ denotes the RF signal action matrix $\partial(\epsilon {\bf C}_1)/\partial [\epsilon {\bf H}_1]_{k\ell}$, ${\bf C}_v$ corresponds to the Doppler-$v$-related coefficient that acts as a thermal correction to the zero-velocity evolution matrix ${\bf C}_0$, and $\bar{\bf z}_v$ is the steady-state solution of the velocity class $v$ that satisfies $({\bf C}_0+(v/\sigma_v){\bf C}_v)\bar{\bf z}_v=-(1/2){\bf w}_0$. To be specific, the definition of ${\bf C}_v$ is given by 
\begin{equation}
    {\bf Q}\T {\bf A}_v{\bf Q} = \begin{bmatrix} 0 & {\bf 0}_{1\times 15} \\ {\bf 0}_{15\times 1} & {\bf C}_v\end{bmatrix},
\end{equation}
where the matrix ${\bf A}_v\in\mathbb{C}^{16\times 16}$ is defined as 
\begin{equation}
\begin{aligned}
    {\bf A}_v &= -\ri k_p\sigma_v \left({\bf I}_4\otimes {\bf D}_p - {\bf D}_p\otimes {\bf I}_4\right) \\ 
    &~+\ri k_c\sigma_v \left({\bf I}_4\otimes {\bf D}_c - {\bf D}_c\otimes {\bf I}_4\right), 
\end{aligned}
\end{equation}
${\bf D}_p={\rm diag}([0, -1, -1, -1])$, and ${\bf D}_c={\rm diag}([0, 0, -1, -1])$.  }

\begin{theorem}[Analytical dynamic response] \label{theorem1}
\red{
For any given $s\in\mathbb{C}$, the Doppler-averaged transfer function $T_{k\ell}^{\rm D}(s)$ equals the 2nd component of the following expression 
\begin{equation}
\begin{aligned}
    &  {\bf Q}\begin{bmatrix}{\bf 0} \\ {\bf I}_{15}\end{bmatrix} \sum_{m,n=1}^{15} \mathbb{E}_X\left[\frac{1}{(1-\lambda_m(s)X)(1-\lambda_n(0)X)}\right] \\ &~~~~\times {\bf s}_m(s){\bf t}_m\T(s) (s{\bf I}-{\bf C}_0)^{-1} {\bf F}_{k\ell} {\bf s}_n(0) {\bf t}\T_n(0)\bar{\bf z}_0,
\end{aligned} \label{eqn:Tkl_analytical}
\end{equation}
where $(\lambda_j(s), {\bf s}_j(s), {\bf t}_j(s))$ are respectively the $j$-th eigenvalue, right eigenvector, and left eigenvector of the matrix $(s{\bf I}-{\bf C}_0)^{-1}{\bf C}_v$. 
The expectation $\mathbb{E}_X[\cdot]$ is taken with respect to a standard Gaussian distribution $X\sim\mathcal{N}(0,1)$, which can be analytically expressed as 
\begin{equation}
\begin{aligned}
    &\mathbb{E}_X\left[\frac{1}{(1-\lambda_m(s)X)(1-\lambda_n(0)X)}\right] \\ & = \begin{cases}\frac{J(\lambda_m^{-1}(s))-J(\lambda_n^{-1}(0))}{\lambda_m(s)-\lambda_n(0)} & {\rm if} \,\lambda_m(s),\lambda_n(0) \neq 0 \\ \lambda_m^{-1}(s)J(\lambda_m^{-1}(s)) & {\rm if}\,\lambda_n(0) =0,\lambda_m(s)\neq 0 \\ \lambda_n^{-1}(0)J(\lambda_n^{-1}(0)) & {\rm if}\,\lambda_m(s)=0,\lambda_n(0) \neq 0. \end{cases}
\end{aligned}
\end{equation}
The special function $J(z)$ is defined for every $z\in\mathbb{C}$ as 
\begin{equation}
    J(z) = \mathcal{N}(z|0,1)\left(\pi {\rm erfi}\left(z/\sqrt{2}\right) -i\pi {\rm sgn}\,\Im{z}\right).
    \label{eqn:J(z)_analytical_expression}
\end{equation}
}
\end{theorem}

\begin{IEEEproof}
See {\bf Appendix~\ref{appendix_1}}. 
\end{IEEEproof}

Similar to $T_{k\ell}^{\rm D}$, the in-phase gain $G_{\rm I}$ and the quadrature gain $G_{\rm Q}$ can be upgraded to $G_{\rm I}^{\rm D}$ and $G_{\rm Q}^{\rm D}$ by applying {\bf Theorem~\ref{theorem1}} to~\eqref{eqn:GI_GQ}. 

\red{
\begin{remark}
    Following a similar approach as in~\cite{tang2025noise}, the transfer functions $T_{k\ell}^{\rm D}(s)$ can also be used to compute the effects of laser amplitude noises and phase noises on the atomic response. For the probe laser, $T_{12}^{\rm D}$ and $T_{21}^{\rm D}$ are employed; for the control laser, $T_{23}^{\rm D}$ and $T_{32}^{\rm D}$ are employed. 
\end{remark}
}

\subsection{Quantum transconductance}
In the previous subsection, we have theoretically computed the transfer function from $X(t)$ to $\Delta\rho_{21}(t)$. In this subsection, we introduce a novel concept of ``quantum transconductance'' to fully describe the transient response from the input E-field $E_{\rm sig}(t)$ to the output signal photocurrent $\Delta I_{\rm sig}(t)$ after photo-electric conversion of the photodiode. This quantum transconductance is analogous to the transconductance of an RF transistor that is key to an RF low-noise amplifier (LNA). 

We define the time-domain quantum transconductance $g_q(x, \tau)$ to be the impulse response from the in-phase component of the signal E-field $\Re{E_{\rm sig}(t)}\,{\rm [V/m]}$ to the output photocurrent change $\Delta{I}_{\rm sig}(t)\,{\rm [A]}$ per unit electro-atomic interaction length $\diff x\,{\rm [m]}$. Note that this time-domain quantum transconductance have a dimension of $\rm [S\cdot Hz]$. By definition of the quantum transconductance, we have 
\begin{equation}
    \Delta I_{\rm sig}(t) = \int_0^L \diff x \int_0^{t} \diff \tau g_q(x, \tau)\Re{E_{\rm sig}}(x, t-\tau),
    \label{eqn:quantum_trans_conductance}
\end{equation}
where the $s$-domain quantum transconductance can be obtained by taking the Laplace transform of both sides, which yields 
\begin{equation}
    \Delta{I}_{\rm sig}(s) = \int_0^L g_q(x,s)E_{\rm sig, 0}I(x,s) \diff x. 
    \label{eqn:quantum_trans_conductance_s_domain}
\end{equation}
This $s$-domain quantum transconductance $g_q(x,s)$ does have a dimension of conductance $[\rm S]$. To compute $g_q(x,s)$, we note that the variations in the probe transmission~\eqref{eqn:probe_transmission} is determined by the variation in the attenuation constant $\Delta\alpha(x,t)$, which is further determined by the variations in the imaginary part of $\Delta\rho_{21}(x,t)$. Thus, by applying all the relationships~\eqref{eqn:probe_transmission},~\eqref{eqn:attenuation_constant}, and~\eqref{eqn:GI_GQ}, the quantum transconductance is expressed in the frequency domain $s=\ri\omega$ as 
\begin{equation}
\begin{aligned}
    g_q(x, \ri\omega) &:= \frac{\diff^2 (\Delta{I}_{\rm sig}) }{\diff E_{\rm sig}\diff x} \\
    &= \frac{q_e\eta}{\hbar\omega_p}\cdot\bar{P}\cdot\frac{2k_p N_0\mu_{12}^2}{\epsilon_0\hbar\Omega_p(x)}\cdot G_{\rm I,2}(x,\ri\omega) \cdot \frac{\mu_{\rm RF}}{\hbar} \\ 
    & \approx \bar{I}_{\rm ph} \cdot \frac{2k_p N_0\mu_{12}^2}{\epsilon_0\hbar\Omega_p(x)} \cdot G_{\rm I,2} (x, \ri\omega)\cdot \frac{\mu_{\rm RF}}{\hbar},  
\end{aligned} \label{eqn:def_quantum_trans_conductance}
\end{equation}
where $\bar{I}_{\rm ph}$ is the DC photocurrent when the signal field is not applied. One can also relate $\Delta {I}_{\rm sig}(\ri\omega)$ to the Fourier transform of the input signal $E_{\rm sig}(x,\ri\omega)$ via 
\begin{equation}
    \Delta{I}_{\rm sig}(\ri\omega)=\int_0^L g_q(x,\ri\omega)\frac{E_{\rm sig}(x,\ri\omega)+E_{\rm sig}^{*} (x,-\ri\omega)}{2}\diff x. \label{eqn:photocurrent_determination}
\end{equation}
From this equation, we can see that the image frequency at $-\ri\omega$ also contributes to the probe response. We note that, in the case where $E_{\rm sig}(x,\ri\omega)$ is completely within the upper sideband, the expression of the photocurrent response $\Delta I_{\rm sig}(t)$ can be simplified to be $\Delta I_{\rm sig, a}(t) = \int_0^L \left(g_q(x,\tau){*}E_{\rm sig}(x,\tau)\right)\vert_{\tau=t}\diff x$, where $*$ denotes convolution.  

Generally, $g_q$ depends on $x$ through the probe Rabi frequency $\Omega_p(x)$. This dependence is caused by the reduction of $\Omega_p(x)$ as a function of $x$ caused by probe attenuation. For small vapor cells with short interaction length $L$, this dependence is somehow weakened. 

\begin{remark}
The transient probe response is mainly described by $G_{{\rm I},2}$ in~\eqref{eqn:re_im_transfer_functions}, whose constant multiple is defined as the quantum transconductance~\eqref{eqn:def_quantum_trans_conductance}. The quantum transconductance for other components can be similarly defined by applying the factor $2k_p N_0\mu_{12}^2\bar{I}_{\rm ph}/(\epsilon_0\hbar\Omega_p)$ to the other three transfer functions in~\eqref{eqn:re_im_transfer_functions}, but they are significantly smaller than $G_{{\rm I},2}$, as will be verified in the numerical results section. 
\end{remark}


\begin{table*}[ht]
    \captionred
    \centering
    \begin{threeparttable} 
        \caption{Proposed dynamic signal models for Rydberg atomic receivers} \label{tab:model_summary}
        \vspace{-3pt}
        \setstretch{1.02}
        \begin{tabular}{|c|c|c|c|c|}
            \hline  
            \multirow{2}{*}{\bf Model category}& \multicolumn{2}{c|}{\bf Time-domain} & \multicolumn{2}{c|}{\bf Frequency-domain}   \\ 
            \cline{2-5} 
                        & with Doppler & without Doppler & with Doppler & without Doppler \\ 
            \hline 
            \multirow{2}{*}{\bf Numerical}   & R-K method  & \multirow{2}{*}{R-K method on~\eqref{eqn:master_equation}} & \multirow{2}{*}{Thermal average on~\eqref{eqn:Tkl_formula}}            & \multirow{2}{*}{\eqref{eqn:Tkl_formula}}               \\ 
             & thermal average for~\eqref{eqn:doppler_average_rho}  &  &  &  \\  
            \hline 
            \multirow{2}{*}{\bf Analytical}  & \multirow{2}{*}{N/A}        & $g_q(\tau)$ computed from             &  $g_q^{\rm D}(\ri\omega)$  computed from   & $g_q(\ri\omega)$ computed from        \\  
             & & ILT of~\eqref{eqn:Tkl_formula} & \eqref{eqn:Tkl_analytical} and~\eqref{eqn:def_quantum_trans_conductance} & \eqref{eqn:Tkl_formula} and~\eqref{eqn:def_quantum_trans_conductance} \\ 
            \hline
        \end{tabular}
        {\footnotesize ILT: inverse Laplace transform; R-K: Runge-Kutta.  }
    \end{threeparttable}
\end{table*}

The proposed dynamic signal models for Rydberg atomic receivers are summarized in~{\bf Table~\ref{tab:model_summary}}. Note that the dynamic signal models are applicable in the {\it near-equilibrium} regime, where small-signal perturbation condition is satisfied.

\subsection{Alternative definition of the Rydberg intrinsic gain} \label{sec3-5}
The authors of~\cite{jing2020atomic,chen2025harnessing} define the response of the Rydberg atomic receiver to a single-tone input to be 
\begin{equation}
    P(t) = \bar{P}+ \kappa|\Omega_{\rm sig}|\cos(2\pi f_{\rm IF}t+\theta), \label{eqn:P(t)_kappa}
\end{equation}
where $\kappa$ is the intrinsic gain of the Rydberg system with unit ${\rm [W\,Hz^{-1}]}$, and $\theta$ is the phase of the signal field. Here we extend this intrinsic gain $\kappa$ into the frequency-dependent version $\kappa(\ri\omega)$. Assuming a uniform signal field across the quantum aperture $[0,L]$, the new definition of $\kappa$ is naturally related to $G_{{\rm I},2}(x,\ri \omega)$ by 
\begin{equation}
    \kappa(\ri\omega) = \bar{P} \int_0^L \frac{2k_pN_0\mu_{12}^2}{\epsilon_0\hbar\Omega_p(x)}G_{{\rm I},2}(x,\ri\omega)\diff x, 
\end{equation}
where $\omega = 2\pi f_{\rm IF}$. Note that for $\omega=0$, the definition and computation formulas of the DC intrinsic gain $\kappa(\ri 0)$ are in alignment with the theoretical work in~\cite{chen2025harnessing}. A straightforward numerical computation of the DC gain $\kappa(\ri0)$ yields the result of $-8.67\times 10^{-13}\,{\rm [W/Hz]}$, which is aligned with the measurement result in~\cite[Table.~1]{feng2023research}. \red{The benefit of introducing a fully frequency-domain $\kappa(\ri\omega)$ is to capture the atomic system response with multiple (often infinitely many) sinusoidal E-field inputs. In this case, the probe response $P(t)$ in~\eqref{eqn:P(t)_kappa} can be extended to be }
\red{
\begin{equation}
    P(t) = \bar{P} + \frac{1}{2\pi}\Re{\int_{0}^{\infty} {\kappa}(\ri\omega)\Omega_{\rm sig}(\omega)e^{\ri\omega t}\diff \omega},
\end{equation}
where $\Omega_{\rm sig}(\omega)$ is the Fourier transform of $\Omega_{\rm sig}(t)$. 
}

\red{Furthermore, the frequency-dependent intrinsic gain is related to the quantum transconductance $g_q(x, \ri\omega)$ via 
\begin{equation}
    \kappa(\ri\omega) = \frac{\hbar\omega_p}{q_e \eta}\cdot\frac{\hbar}{\mu_{\rm RF}} \int_0^L g_q(x, \ri\omega)\diff x. 
\end{equation} } 
Compared to the frequency-dependent intrinsic gain $\kappa(\ri\omega)$, the proposed quantum transconductance $g_q(x,\ri\omega)$ is $x$-dependent. Since the blackbody radiation (BBR) exhibits strong spatial correlation with coherence length $\sim\lambda/2$~\cite{bertilone1997spatial,carminati1999near}, this coordinate-dependent property of $g_q(x,\ri\omega)$ enables the subsequent blackbody radiation noise analysis of the atomic receiver.

\section{Noise Models}\label{sec4}
Noises that affect Rydberg atomic receivers can be generally divided into two categories: the external noises, and the internal noises. While internal noises can be actively suppressed by optimized designs of the receiver, external noises pose intrinsic limits to the atomic receiver, since the receivers cannot distinguish in-band external noises from the desired signals. In this section, we study how the external in-band blackbody radiation noises affect the response of the atomic receiver. 

\subsection{Properties of the blackbody radiation}
Blackbody radiation is the wide-spectrum radiation emitted by any object above absolute zero temperature. In the normal temperature environment of $T=300\,{\rm K}$, blackbody radiation contributes to most part of the electronic noise observed by classical RF receivers, since it directly interacts with the RF antenna and creates an equivalent noise source in the first stage of classical RF signal processing pipeline before the low-noise amplifier (LNA). In an isotropic blackbody radiation field of temperature $T$, the spectral radiance is given by
\begin{equation}
    B_\nu(T) = \frac{2\nu^2}{c^2}\cdot\frac{h\nu}{\exp(h\nu/(k_{\rm B}T))-1} \quad {\rm [W\,Hz^{-1}\,m^{-2}\,{rad}^{-1}]},
\end{equation}
where $\nu$ is the center frequency that the blackbody radiation measurement is performed. At room temperature, this spectral radiance is well-approximated by $2\nu^2k_{\rm B}T/c^2$. To compute the atomic response to the blackbody radiation, we need to first compute the per-Hertz electric field phasor as the E-field input $E_{\rm sig}(x, \ri\omega)$ in~\eqref{eqn:quantum_trans_conductance}. By noticing the plane-wave power flux formula $S=|{E}_{\rm sig}|^2/(2\eta_0)$, we write the unit-solid angle unit-frequency blackbody radiation field at point ${\bf x}$ along incidence direction $\hat{\bf r}$ to be 
\begin{equation}
    {\rm d}{\bf E}_{\nu}({\bf x},\hat{\bf r}) = \sqrt{\eta_0 B_\nu(T,\hat{\bf r})} e^{\ri k_0\hat{\bf r}\cdot{\bf x}}(W_1(\hat{\bf r})\hat{\bf t}_1 + W_2(\hat{\bf r})\hat{\bf t}_2){\rm d}S(\hat{\bf r}),
\end{equation}
where $\hat{\bf r}, \hat{\bf t}_1, \hat{\bf t}_2$ constitutes an orthonormal basis of $\mathbb{R}^3$,  $W_1, W_2$ are two independent complex white Gaussian random fields of unit variance on the unit sphere $S^2$, and ${\rm d}S(\hat{\bf r})$ is the standard spherical measure that integrates to $4\pi$. The total strength of the omni-directional noise field is given by ${\bf E}_\nu({\bf x})=\int_{4\pi} \diff {\bf E}_\nu({\bf x}, \hat{\bf r})$. Note that this total field further induces a spatial correlation function given by 
\begin{equation}
    {\bf R}_{\bf E_\nu}({\bf x}, {\bf x}') := \mathbb{E} \left[{\bf E}_\nu({\bf x}) {\bf E}_\nu\H({\bf x}')\right] \quad[{\rm V^2\,m^{-2}\,Hz^{-1}}], \label{eqn:noise_spatial_correlation}
\end{equation}
where $\mathbb{E}[\cdot]$ denotes ensemble average. This quantity can be analytically computed to be 
\begin{equation}
\begin{aligned}
    {\bf R}_{{\bf E}_\nu}({\bf x}, {\bf x}') &= \pi\eta_0B_\nu(T) \\
    &\times \left[(f_0(\beta)+f_2(\beta)){\bf I}_3+(f_0(\beta)-3f_2(\beta))\hat{\bf r}\hat{\bf r}^\mathsf{T}\right],
\end{aligned}\label{eqn:spatial_correlation_closed_form}
\end{equation}
where $f_n(\beta) := \int_{-1}^{1}x^ne^{\ri \beta x}\diff x$, $\beta = k_{\rm LO} r$, $r=\|{\bf x} - {\bf x'}\|$, and $\hat{\bf r} = ({\bf x}-{\bf x'})/r$, and . The same spatial correlation formula of blackbody radiation fields can be found in~\cite{zhu2025electromagnetic,bertilone1997spatial,wan2023mutual}. 
Intuitively, if the blackbody radiation field correlation is stronger, then it will add up coherently from $x=0$ to $x=L$, resulting in a stronger influence on the output of Rydberg atomic receivers. This intuition will be justified by quantitative computations in the next subsection.

\red{
\begin{remark}
    The blackbody radiation is fundamental to Rydberg atomic receivers. In classical antenna theory, blackbody radiation (BBR) interacts with the receive antennas, creating an equivalent noisy wave source of single-sided power spectral density $k_{\rm B}T$ at the antenna port, which is also known as the thermal noise of antenna~\cite{kong1975theory}. This noise is fundamentally determined by the thermal environment, which cannot be reduced by any experimental technique such as cooling the antenna. Since the physical nature of the blackbody radiation is a wideband omnidirectional EM field which covers the incident signal band, it is impossible for the Rydberg atomic receiver to distinguish between the noise field and the signal field. Thus, unlike the atomic transit noise and laser noise, the BBR noise is not reducible by experimental improvements.  
\end{remark}
}

\subsection{Atomic response to the blackbody radiation}
\begin{figure}[t]
    \centering
    \includegraphics[width=\figsize\linewidth]{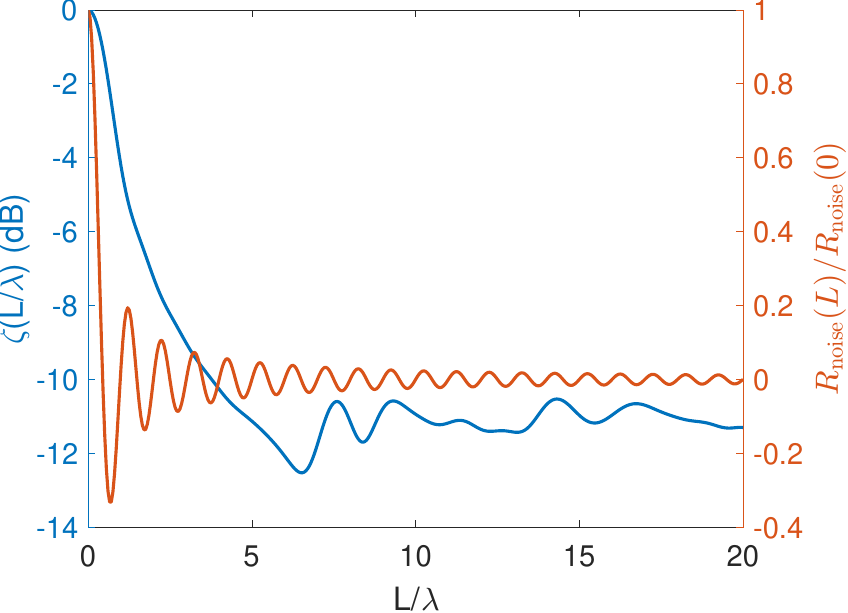}
    \caption{Coherence factor $\zeta(\ell)$ and normalized noise correlation $R_{\rm n}(\ell, \ri 0)$ as a function of normalized interaction length $\ell$. The fluctuation of $\zeta(\ell)$ in the large-$\ell$ regime is caused by the long-tail effect in the noise correlation. }
    \label{fig:zeta_and_R}
\end{figure}

Let $E_{\rm n}(x,\ri\omega)$ be the blackbody radiation noise field phasor with unit $\rm V\,m^{-1}\,Hz^{-1/2}$. The atomic response to this noise, in terms of the power spectral density (PSD) $P_{\Delta{I}_{\rm n,bb}}(\omega)\,[{\rm A^2\,Hz^{-1}}]$ of the noise-induced photocurrent, is defined at IF angular frequency $\omega\neq0$ as 
\begin{equation}
\begin{aligned}
&P_{\Delta{I}_{\rm n,bb}}(\omega):= {\rm PSD}[\Delta I_{\rm bb}(t)](\omega) \\
&= \iint_{[0,L]^2}{\rm d}x_1 {\rm d}x_2 g_q(x_1, \ri\omega)g_q^*(x_2, \ri\omega) \times \\
& \, \mathbb{E}\left[\frac{(E_{\rm n}(x_1, \ri\omega)+E_{\rm n}^*(x_1, -\ri\omega))(E_{\rm n}^*(x_2, \ri\omega)+E_{\rm n}(x_2, -\ri\omega))}{4}\right].  \\
\end{aligned}
\end{equation}
By applying the noise correlation formula $\mathbb{E}[E_{\rm n}(x_1, \ri\omega_1)E_{\rm n}^*(x_2, \ri\omega_2)] = R_{\rm n}(x_1-x_2, \ri\omega_1)\mathbf{1}_{\omega_1=\omega_2}$, we compute the noise current PSD  to be 
\begin{equation}
\begin{aligned}
&P_{\Delta I_{\rm n,bb}}(\omega) \overset{(a)}{\approx} \frac{|g_q(\ri\omega)|^2}{4}\times   \\
& \iint_{[0,L]^2} \left({R_{\rm n}(x_1-x_2,\ri\omega)+R_{\rm n}(x_2-x_1,-\ri\omega)}\right) {\rm d}x_1{\rm d}x_2\\
&\overset{(b)}{\approx} \frac{|Lg_q(\ri\omega)|^2}{2}\cdot\frac{1}{L^2}\int_{-L}^{L}(L-|u|)R_{\rm n}(u,0){\rm d}u, 
\end{aligned}
\end{equation}
where the approximation (a) holds for small electro-optical interaction length $L$, the approximation (b) is justified by approximating $R_{\rm n}(u, \pm\ri\omega)$ with $R_{\rm n}(u,0)$ for small intermediate frequency $|\omega|\ll 2\pi f_{\rm LO}$, and $R_{\rm n}(u, \ri\omega)$ is the $z$-direction noise correlation function at spatial separation $u$, which can be computed by taking the $(3,3)$-th entry in~\eqref{eqn:noise_spatial_correlation}. Since the noise variance is intimately related to the asymptotic rate of BBR coherence decay as a function of $\ell$, we define the coherence factor $\zeta(\ell)$ to be 
\begin{equation}
\begin{aligned}
    \zeta(\ell) &= \frac{1}{L^2R_{\rm n}(0,\ri 0)}\int_{-L}^{L}(L-|u|)R_{\rm n}(u,\ri0){\rm d}u \\
    &= \frac{1}{\ell^2R_{\rm n}(0,\ri 0)}\int_{-\ell}^\ell (\ell - |u'|)R_{\rm n}(\lambda_{\rm LO}u',\ri 0)\diff u',
\end{aligned}
\end{equation}
where $\ell := L/\lambda_{\rm LO}$ is the electro-optical interaction length normalized by LO wavelength. One can check that this coherence factor only depends on $\ell = L/\lambda_{\rm LO}$ instead of $L$, and that it satisfies $0<\zeta(\ell)<1$. When $\ell\to 0$, the spatial correlation becomes almost fully coherent, and thus $\zeta(\ell)$ will be close to $1$. The coherence factor and the normalized noise correlation function are shown in Fig.~\ref{fig:zeta_and_R}. 

With the concept of the spatial coherence factor $\zeta(\ell)$, the PSD of the atomic response current  (variance of the output photoelectric current per unit bandwidth) to the noise field is further simplified to be 
\begin{equation}
    P_{\Delta{I}_{\rm n,bb}}(\omega) = \frac{4\pi}{3}\eta_0B_\nu(T)\zeta\left(\frac{L}{\lambda_{\rm LO}}\right)L^2|g_q(\ri \omega)|^2, 
    \label{eqn:lower_bound_current_variance}
\end{equation}
where $R_{\rm n}(0, \ri\omega)$ is expressed in terms of the spectral radiance $B_\nu(T)$ via~\eqref{eqn:spatial_correlation_closed_form}. Note that the PSD value~\eqref{eqn:lower_bound_current_variance} is only a lower bound in practical Rydberg atomic receivers. In real-world quantum receivers, the internal noises including photon shot noise, atomic collision noise, atomic transition noise, and Johnson-Nyquist noise should also be considered, resulting in a larger noise variance (see Fig.~\ref{fig:Conceptual_RARE} for the noise introduction points in the system). Note that the out-of-band BBR noises that are resonant to other Rydberg-Rydberg transitions have already been considered in the form of dephasing coefficients $\gamma_{3,4}$~\cite{fancher2021rydberg, gong2024rydberg}.  

Assume that the received signal $E_{\rm sig}(\ri\omega)$ completely lies in the upper sideband (USB) with $\omega>0$, such that there is no image frequency interference at $\omega<0$. 
Combining the above BBR noise variance with the signal transfer relationship~\eqref{eqn:quantum_trans_conductance_s_domain}, we compute the best achievable receive signal-to-noise ratio (SNR) at IF angular frequency $\omega>0$ to be 
\begin{equation}
    \mathsf{SNR}_{q}^{\star}(\omega) = \frac{\frac{1}{4}P_{\rm sig}(\omega)}{\frac{4\pi}{3}\eta_0 B_\nu(T)\zeta(L/\lambda_{\rm LO})}, 
    \label{eqn:quantum_receiver_SNR}
\end{equation}
where $P_{\rm sig}(\omega)\,{\rm [V^2\,m^{-2}\,Hz^{-1}]}$ is the double-sided power spectral density of the infinitely long {\it analytic} input signal ${E_{\rm sig}}(t)$, and the blackbody spectral radiance $B_\nu$ is evaluated at $\nu = f_{\rm LO}$. The factor $1/4$ is explained by the selective response to the in-phase component $I(t)$ of the Rydberg atomic receivers and the definition of double-sided PSD.

\subsection{Sensitivity analysis}
Another popular performance measure of the Rydberg atomic receiver is the minimum detectable E-field. By setting the SNR in~\eqref{eqn:quantum_receiver_SNR} to 0\,dB, we get the lower bound of the per-second detectable in-phase E-field strength to be 
\begin{equation}
    E_{\rm I, min} \geq \sqrt{\frac{4\pi}{3}\eta_0 B_\nu(T)\zeta(\ell)} \quad [{\rm V\, m^{-1} \,Hz^{-1/2}}].\label{eqn:blackbody_sensitivity_limit}
\end{equation}
A straightforward computation at $T=300{\rm K}$, $f_{\rm LO} = 6.9458\,{\rm GHz}$ with $\zeta=1$ yields the value of $E_{\rm I, min}\geq 838\,{\rm pV\,cm^{-1}\,{Hz}^{-1/2}}$, while this value can be improved by increasing the length of electro-atomic interaction $L$. This blackbody radiation-determined value is comparable to the quantum projection noise limit (QPNL)-determined sensitivity of $\sim 700\,{\rm pV\,cm^{-1}\,Hz^{-1/2}}$~\cite{jing2020atomic}. Although a measured sensitivity of $\sim 55\,{\rm nV\,cm^{-1}\,Hz^{-1/2}}$~\cite{jing2020atomic} and $10.0\,{\rm nV\,cm^{-1}\,Hz^{-1/2}}$~\cite{tu2024approaching} are reported, which is about 20\,dB worse than state-of-the-art classical electronic receivers~\cite{gong2024rydberg}, we anticipate that the sensitivity of Rydberg atomic receivers will finally approach the lower bound that is jointly determined by~\eqref{eqn:blackbody_sensitivity_limit} and the QPNL sensitivity bound.

\subsection{Noise PSD in the output power of the photodiode} 

\begin{figure*}[t]
    \centering
    \includegraphics[width=\largeFigSize\linewidth]{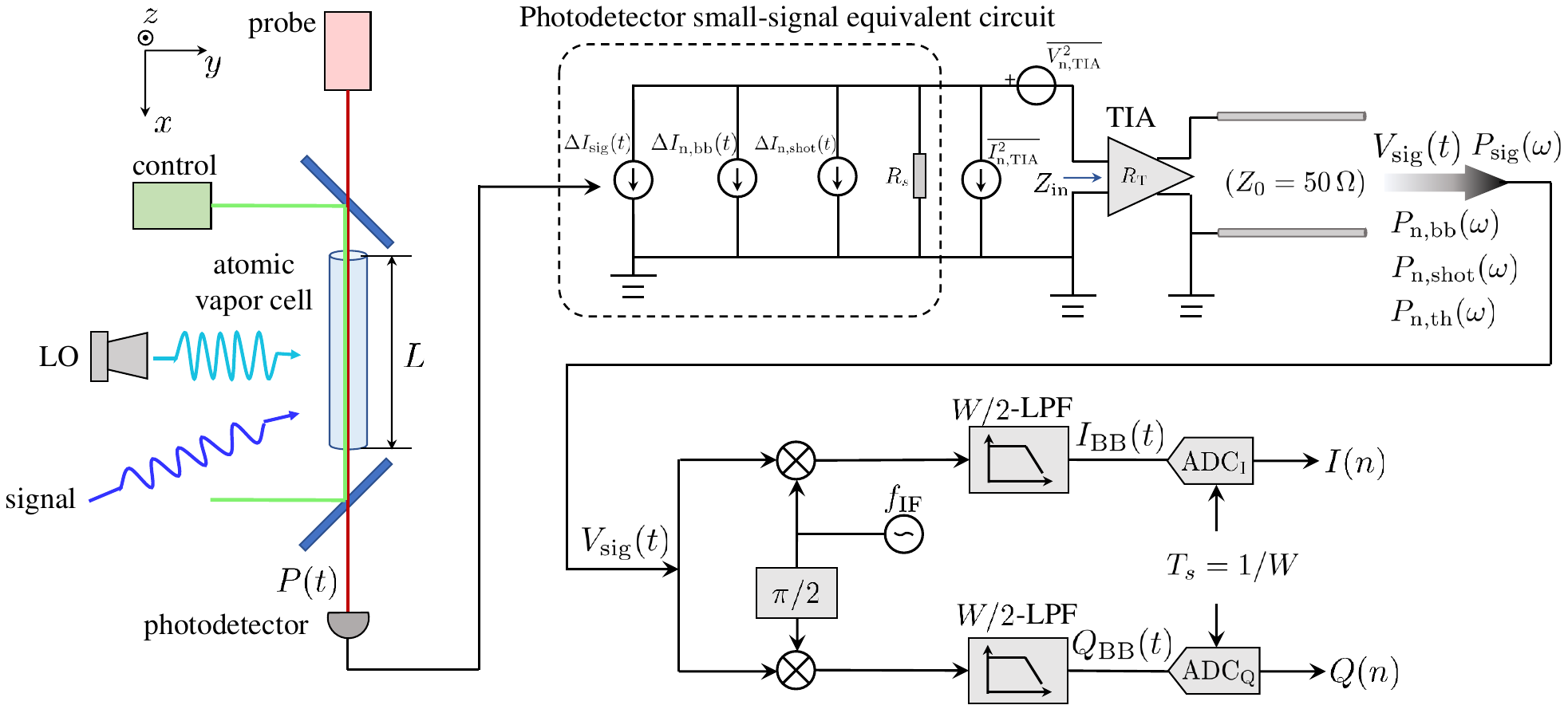}
    \caption{Analog signal processing pipeline for Rydberg atomic receivers: Photo-atomic interaction, transimpedance amplification (TIA), and IQ down-conversion. }
    \label{fig:sp_circuits}
\end{figure*}

In most of the applications, it is more beneficial to know the electronic noise parameters in terms of the PSD of the noise current, since this PSD can be directly measured by the spectrum analyzers. The opto-electronic signal processing pipeline of Rydberg atomic receiver is shown in Fig.~\ref{fig:sp_circuits}.

The small-signal output photocurrent $\Delta I_{\rm ph}(t)$ is expressed in the time domain as  
\begin{equation}
    \Delta I_{\rm ph}(t) = \Delta {I}_{\rm sig}(t) + \Delta {I}_{\rm n,bb}(t) +  \Delta I_{\rm n,R_s}(t)+\Delta I_{\rm n,shot}(t),  
\end{equation}
where $\Delta{I}_{\rm sig}(t)$ is determined by the input signal E-field through~\eqref{eqn:quantum_trans_conductance}, the PSD of $\Delta{I}_{\rm n,bb}(t)$ is lower-bounded by~\eqref{eqn:lower_bound_current_variance}, the PSD of $\Delta I_{\rm n,R_s}(t)$ equals $2k_{\rm B} T/R_{\rm s}$ with $R_{\rm s}$ being the DC-biasing resistance in series with the photodiode, and $\Delta I_{\rm n,shot}(t)$ is the photon shot noise with PSD given by 
\begin{equation}
    {\rm PSD}[\Delta I_{\rm n,shot}(t)](\omega)=q_e \bar{I}_{\rm ph}. 
\end{equation}
Note that the photon shot noise is caused by the discreteness of the Poisson photon arrival process. To convert this current PSD into the output PSD of the transimpedance amplifier (TIA), we assume that the TIA (FEMTO DHPCA-100~\cite{DHPCA100}) has a transimpedance of $R_{\rm T}=10\,{\rm k\Omega}$, an input referred current noise level of $I_{\rm n,TIA}=1.8\,{\rm pA\,Hz^{-1/2}}$, an input referred voltage noise level of $V_{\rm n, TIA}=2.8\,{\rm nV\,Hz^{-1/2}}$, and an input impedance of $Z_{\rm in}=60\,{\rm \Omega}$. The photodetector is assumed to be a reverse-biased photodiode in series of a resistor $R_s=1\,{\rm k\Omega}$ that is connected to the supply voltage. The TIA output is connected to a matched load of $R_{\rm L}=50\,{\Omega}$~\cite{feng2023research}, e.g. a spectrum analyzer. 

With these assumptions, the conversion formula from the photocurrent noise PSD to the TIA output noise PSD given by 
\begin{equation}
    P_{\rm n,out}(\omega) = {\rm PSD}[\Delta I](\omega)\times \frac{\left(R_{\rm T}K_c\right)^2}{R_{\rm L}} \label{eqn:currentPSD2powerPSD}
\end{equation}
where $P_{\rm n,out}(\omega)$ is the double-sided output noise power spectral density of the TIA measured in $\rm W/Hz$, $K_c=R_s/(R_s+Z_{\rm in})$ is the current divider coefficient at the TIA input, and ${\rm PSD}[\Delta I](\omega)$ is in $\rm A^2/Hz$. Note that ${\rm PSD}[\Delta I](\omega)$ can represent both the BBR-induced noise and the photon shot noise. Thus, $P_{\rm n, out}$ can represent both $P_{\rm n, bb}$ and $P_{\rm n, shot}$. Moreover, the circuit total thermal noise (Johnson-Nyquist noise and the amplifier noise) PSD can be computed by 
\begin{equation}
    P_{\rm n, th}(\omega) =\frac{R_{\rm T}^2}{2R_{\rm L}}\left( I_{\rm n, TIA}^2K_c^2 +\frac{V_{\rm n, TIA}^2}{(R_s+Z_{\rm in})^2}+\frac{4k_{\rm B}T}{R_s}\right) \\
\end{equation}
which is also in unit $\rm W/Hz$. 



\subsection{Noise factors}
\begin{figure}[t]
    \centering
    \includegraphics[width=\figsize\linewidth]{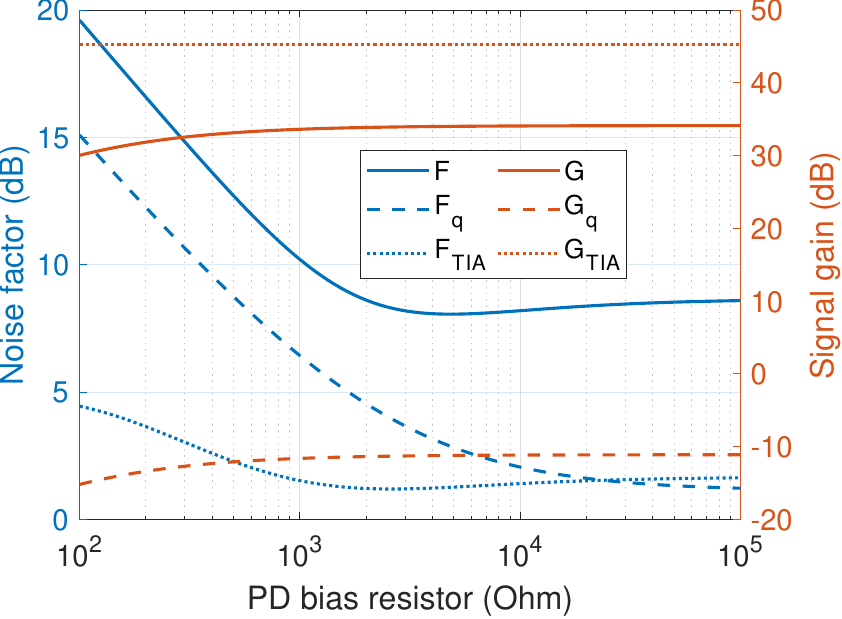}
    \caption{Noise factor and power gain of the quantum LNA (qLNA) and the transimpedance amplifier (TIA) as a function of the PD bias resistor $R_s$. }
    \label{fig:nf}
\end{figure}

To enable a fair comparison between classical electronic receivers and quantum receivers, in this subsection, we compute the noise factors of quantum receivers by applying the Friis formulas for noise. Following the concept of quantum transconductance $g_q$, we treat the quantum receiver and the photodiode as a quantum low-noise amplifier (qLNA) with noise factor 
\begin{equation}
    F_{q} = \frac{\mathsf{SNR}_{\rm in}}{\mathsf{SNR}_{\rm out}} = \frac{A_{\rm eq}|E_{\rm sig}|^2/(2\eta_0 k_{\rm B}T)}{(Lg_q|E_{\rm sig}|)^2/4/{\rm PSD}[{\Delta I_{\rm n}}]}, 
\end{equation}
where $A_{\rm eq}=3\lambda^2/(8\pi)$ is the equivalent aperture of a classical dipole antenna, and the output noise ${\rm PSD}[{\Delta I_{\rm n}}]$ contains the current PSD of BBR noise , photon shot noise, thermal noise of $R_s$, \red{and the relative intensity noise (RIN) of the probe laser is set to be $\bar{I}_{\rm ph}^2\times {\rm PSD}_{\rm RIN}$. Note that this is because the RIN noise does not change after passing the photodiode. Thus, the output current noise PSD of the laser RIN can be computed as 
\begin{equation}
    {\rm PSD}[\Delta I_{\rm n, RIN}](\omega) = (\bar{I}_{\rm ph})^2\cdot {\rm PSD}_{\rm RIN}(\omega). 
\end{equation}
In the numerical simulations, we set ${\rm PSD}_{\rm RIN}=-140\,{\rm dBc/Hz}$. }
The power gain of the qLNA is given by 
\begin{equation}
    G_{q} = \frac{(Lg_q |E_{\rm sig}|K_c)^2 Z_{\rm in}/2}{|E_{\rm sig}|^2A_{\rm eq}/(2\eta_0)}.
\end{equation}
The noise factor and gain of the electronic TIA is given by 
\begin{equation}
    F_{\rm TIA} = 1+\frac{(I_{\rm n,TIA}K_c)^2/2+(V_{\rm n,TIA}/(R_s+Z_{\rm in}))^2/2}{{\rm PSD}[{\Delta I_n}]}, 
\end{equation}
and the TIA power gain is 
\begin{equation}
    G_{\rm TIA} = \frac{R_{\rm T}^2}{Z_{\rm in}R_{\rm L}}.  
\end{equation}
By the Friis formulas for noise, the total power gain is $G = G_q G_{\rm TIA}$, and the total noise factor is $F=F_q+(F_{\rm TIA}-1)/G_q$. The numerical results are shown in \red{Fig.~\ref{fig:nf}}, where it can be seen that the minimum total noise factor of the quantum receiver is $F=8.1\,{\rm dB}$ achieved at $R_s=4\,{\rm k\Omega}$.  

\section{Simulation Results} \label{sec5}
In this section, we present the simulation results of a Rydberg atomic receiver. The atomic receiver equips a cesium-133 atomic vapor cell of length $L=2\,{\rm cm}$, where the vapor cell is illuminated by a probe light at 852\,nm with a probe power of $P_0=29.8\,{\rm \mu W}$ and a probe Rabi frequency of $\Omega_p=2\pi\times8.08\,{\rm MHz}$. The counter-propagating control light at 510\,nm is assumed to have a Rabi frequency of $\Omega_c=2\pi\times 2.05\,{\rm MHz}$~\cite{zhang2019detuning}. The value of the decay rates $\gamma_{i}$ and the other parameters are aligned with~\cite{jing2020atomic}. The quantum efficiency of the photodetector is set to be $\eta = 0.8$. All noises are evaluated at $T=300\,{\rm K}$.

\subsection{Dynamic results in the time domain}
We first check the correctness of the small-signal transfer function~\eqref{eqn:re_im_transfer_functions} by simulating a single-carrier Rydberg atomic communication system. The simulated probe response of this atomic receiver is compared to that of the theoretically predicted response~\eqref{eqn:re_im_transfer_functions}. The simulated probe response is obtained by applying an order-4 Runge-Kutta method to numerically solve the master equation~\eqref{eqn:master_equation}. For the theoretical method, the transfer function is first calculated by evaluating~\eqref{eqn:Tkl_formula} and~\eqref{eqn:re_im_transfer_functions}, and then the transfer function is applied to the received E-field signal $E_{\rm sig}(t)$ to theoretically compute the signal contained in $\Im{\rho_{21}}(t)$. In both simulation and theoretical prediction, the probe response is computed by applying~\eqref{eqn:probe_transmission}.  

\begin{table}[t]
    \centering
    \begin{threeparttable} 
        \caption{Communication System Parameters} \label{tab:CommSyst_Parameter}
        \vspace{-3pt}
        \setstretch{1.07}
        \begin{tabular}{|l|c|}
            \hline 
            Item                        & Value           \\ 
            \hline
            Carrier frequency           & 6.9458 GHz      \\ 
            \hline
            IF frequency $f_{\rm IF}$   & 150 kHz         \\
            \hline
            LO intensity $E_{\rm LO}$   & 0.04 V/m        \\ 
            \hline 
            BS Tx power                 & 10 dBm          \\ 
            \hline 
            BS Antenna gain             & 5 dB            \\ 
            \hline
            Classical UE antenna gain   & 1.76\,dB        \\ 
            \hline 
            BS-UE distance              & 200 m          \\ 
            \hline
            Symbol period               & 10 us           \\ 
            \hline 
            Runge-Kutta step size       & 1 ns            \\
            \hline
        \end{tabular}
    \end{threeparttable}
\end{table}


\begin{figure}[t]
    \centering
    \includegraphics[width=\figsize\linewidth]{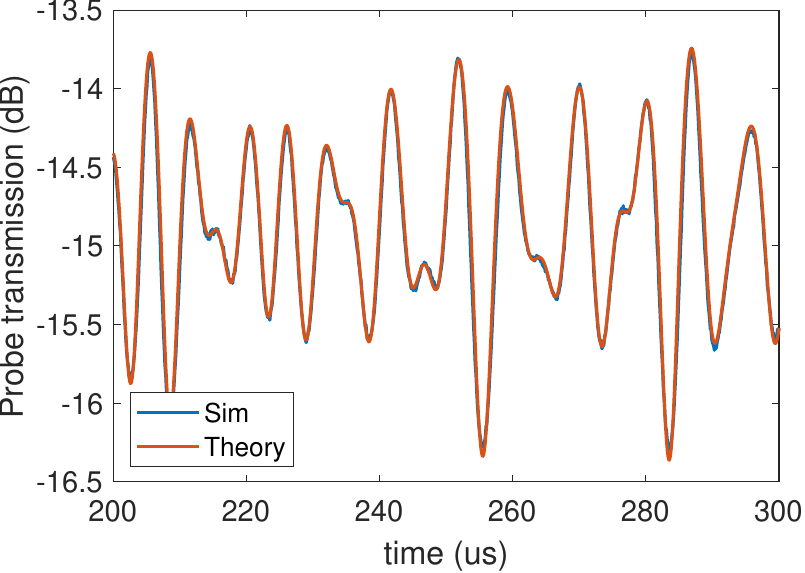}
    \caption{Simulated and theoretically predicted probe responses to a single-carrier modulated signal ($T=0\,{\rm K}$). }
    \label{fig:waveform_fine}
\end{figure}

\red{Figure~\ref{fig:waveform_fine}} shows the simulated and theoretical dynamic response of the Rydberg atomic receiver to an information-carrying single-carrier communication signal. It can be seen that the dynamic component of the theoretical time-domain response well-fits the simulated response. 

\begin{figure}[t]
    \centering
    \includegraphics[width=\figsize\linewidth]{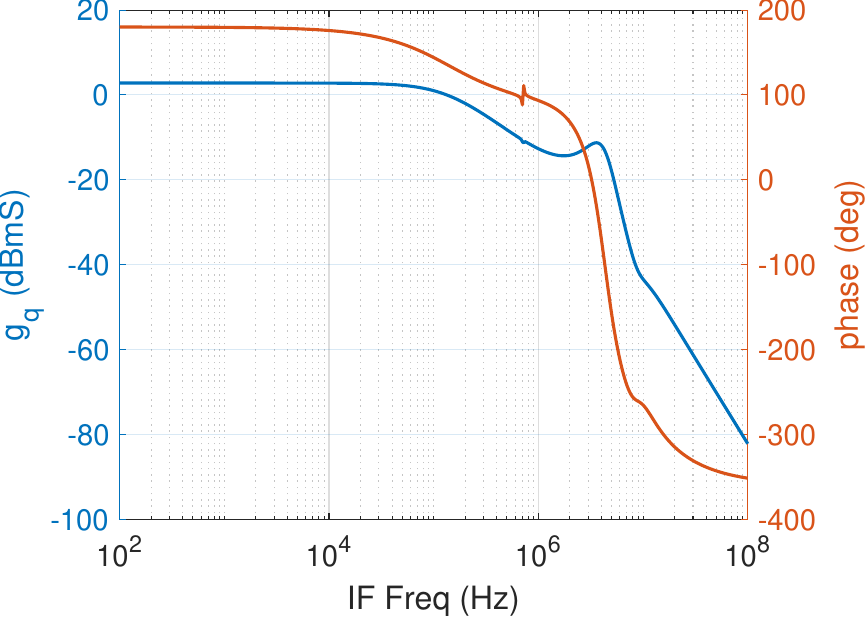}
    \caption{Amplitude and phase responses of the quantum transconductance $g_{q}(\ri\omega)$, evaluated at $T=0\,{\rm K}$. }
    \label{fig:gq}
\end{figure}

\subsection{Time-frequency results of the quantum transconductance}
To better capture the input-output transition relationship of the Rydberg atomic receiver, in this subsection, we numerically compute the quantum transconductance in the time-frequency domain. \red{Figure~\ref{fig:gq}} presents the amplitude and phase response of the quantum transconductance $g_q$ as a function of the IF frequency $f_{\rm IF}$. 

\begin{figure}
    \centering
    \includegraphics[width=\figsize\linewidth]{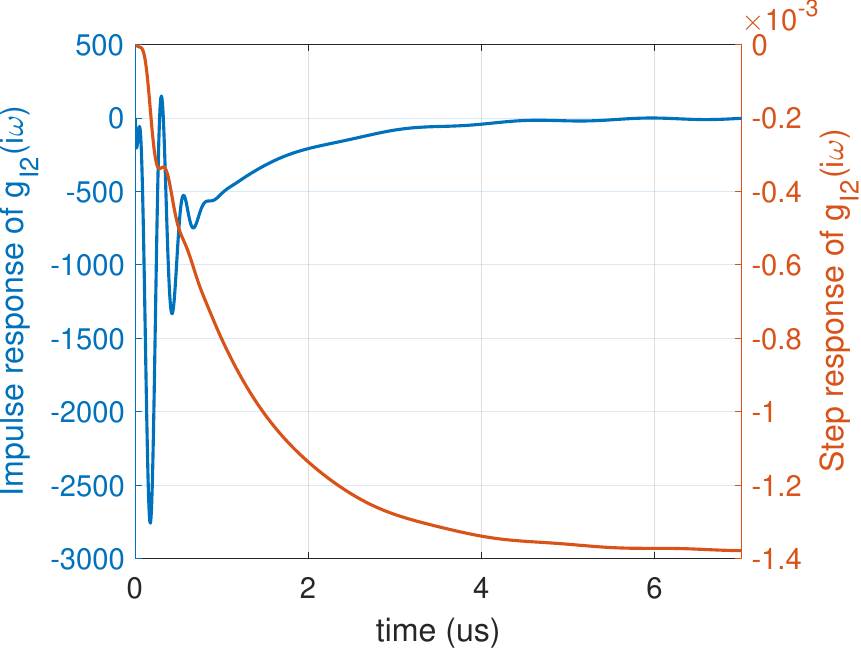}
    \caption{Impulse response $g_q(x,\tau)$ and the corresponding step response $\int_0^t g_q(x,\tau)\diff\tau$ of the quantum transconductance as a function of time, evaluated at $T=0\,{\rm K}$. }
    \label{fig:imp_step_response}
\end{figure}

\red{Figure~\ref{fig:imp_step_response}} shows the impulse response and the step response of the quantum transconductance at zero temperature. \red{It can be seen that the oscillatory behavior in the atomic response~\cite{wu2024based,sapiro2020time} has already been captured by the impulse response $g_q(\tau)$ of the proposed quantum transconductance.} A rising time of $t_r=2.45\,{\rm \mu s}$ can be achieved, which corresponds to a 3\,dB bandwidth of ${\rm BW}_{\rm 3dB}=0.35/t_r\approx 0.14\,{\rm MHz}$, which is in agreement with the results of frequency-domain transconductance $g_q(\ri\omega)$. The theoretical single-sided instantaneous bandwidth value of about $0.1\,{\rm MHz}$ is also supported by experimental works~\cite{anderson2020atomic,deb2018radio,li2022rydberg}.

\begin{figure}
    \centering
    \includegraphics[width=\figsize\linewidth]{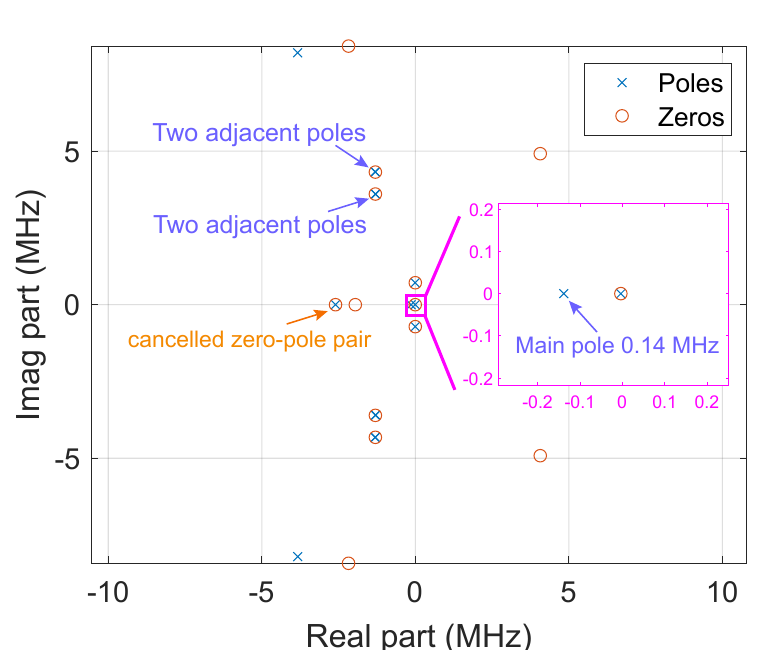}
    \caption{Pole-zero map of the quantum transconductance $g_q(s)$, evaluated at $T=0\,{\rm K}$. }
    \label{fig:pzfig}
\end{figure}

Figure~\ref{fig:pzfig} shows the pole-zero map of the zero-temperature quantum transconductance $g_q(s)$. The quantum dynamic response is fully described by 15 poles, 13 zeros, and a DC gain $g_q(\ri 0)=-1.4\,{\rm mS}$ via the rational fraction factorization 
\begin{equation}
    g_q(s) = g_q(\ri 0) \frac{\prod_{k=1}^{13}(1-\frac{s}{z_k})}{\prod_{\ell=1}^{15}(1-\frac{s}{p_\ell})}. 
\end{equation}
Note that all the poles and zeros are normalized by $2\pi$ to convert into frequency values.  

\begin{figure}
    \centering
    \includegraphics[width=1.0\linewidth]{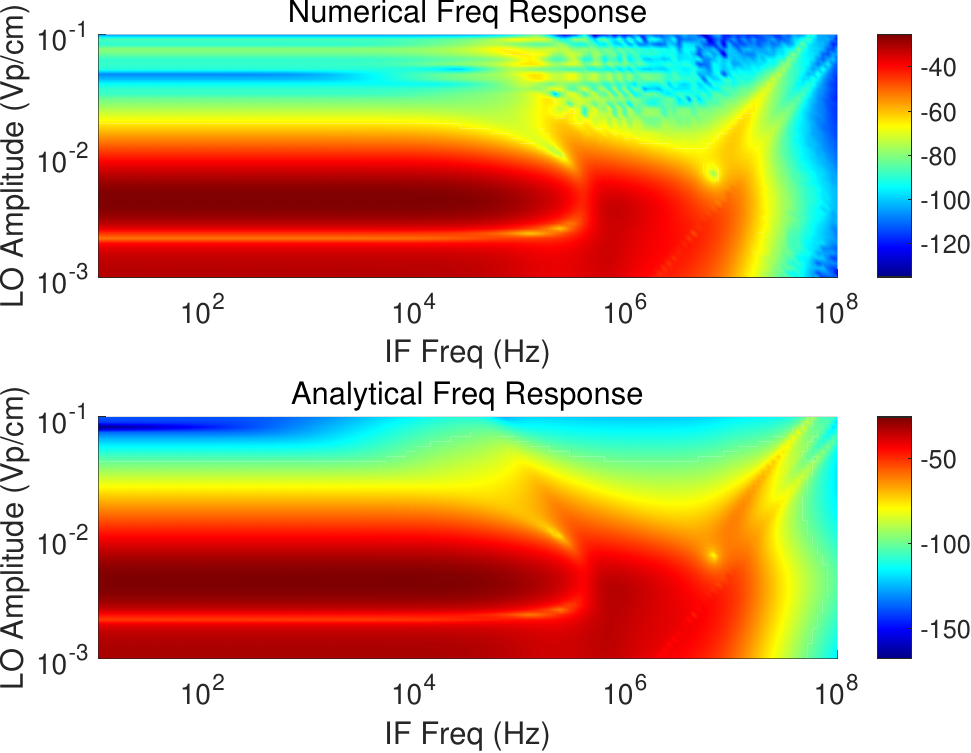}
    \caption{Frequency response $g_q^{\rm D}(\ri\omega)$ evaluated at $T=300\,{\rm K}$. Numerical computation v.s. analytical evaluation. }
    \label{fig:T_NumericalAnalyticalFreqResponse}
\end{figure}

\red{Figure~\ref{fig:T_NumericalAnalyticalFreqResponse} shows the Doppler-aware frequency response $g_q^{\rm D}(\ri\omega)$ at room temperature. The frequency response is computed using two methods: the numerical method and the analytical formula, as is described in {\bf Table~\ref{tab:model_summary}}. The results from two methods coincide well, and the analytical result is smoother due to absence of numerical bins.  }

\subsection{Continuous-time simulation for wireless communications} 
\red{To establish the complete wireless simulation link for the Rydberg atomic receiver, we perform downlink waveform-level simulations by assuming a single-antenna BS serving a UE equipped with a single Rydberg atomic receiver. The channel model is assumed to be line-of-sight (LoS) with no obstacles and time-selective fading. At the Rydberg atomic receiver, the RF signal is first mixed by the LO signal with the Rydberg atoms, creating an optical IF signal with BBR noise PSD described by~\eqref{eqn:lower_bound_current_variance}. Note that the optical response of the atoms are computed from the proposed quantum transconductance model. After opto-electric conversion by the photodiode, this IF signal is amplified with a TIA, and then mixed with two orthogonal IF tones to down-convert the signal to the complex baseband. }

\begin{figure}
    \centering
    \includegraphics[width=\figsize\linewidth]{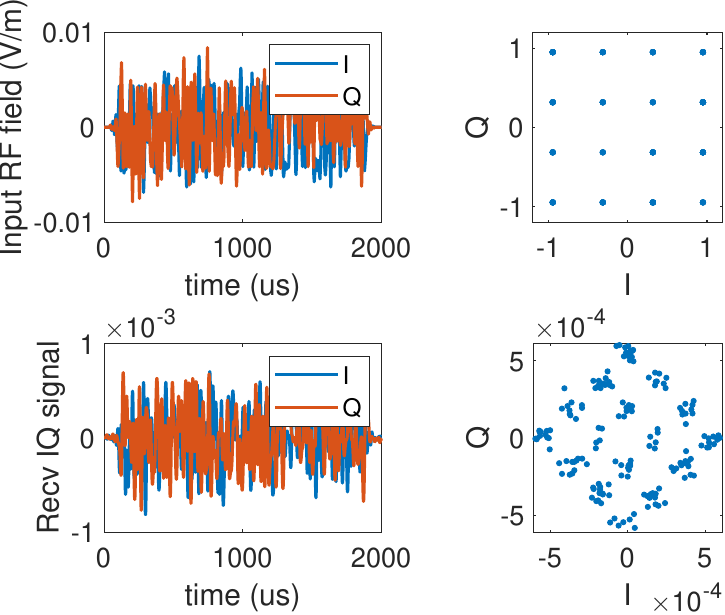}
    \caption{Input noiseless RF waveform (analytic signal at $f_{\rm LO}$) and the received noisy baseband complex signals, as well as their constellation representations. The estimated SNR is $17.71\,{\rm dB}$ from the Rx constellation.  }
    \label{fig:constellations}
\end{figure}

The Tx/Rx waveforms and 16QAM constellations are shown in Fig.~\ref{fig:constellations}, with the imperfections in the received constellation attributed to the TIA electronic thermal noise, the BBR noise, the nonlinear effects of the Rydberg atom receiver, and the imperfect ISI suppression during single-carrier transmission. Detailed inspections of the contributions of these imperfections to the total error vector magnitude (EVM) are beyond the scope of this theoretical paper, and are left for future works. From the results in Fig.~\ref{fig:constellations}, the Rx SNR is estimated to be 17.71\,dB, and thus the equivalent noise PSD of the simulated Rydberg atomic receiver is estimated to be $-152.0\,{\rm dBm/Hz}$, which is about 22\,dB above the thermal noise floor $n_0$. This numerical result approximately coincides with the electric field measurement precision of $10.0\,{\rm nV\,cm^{-1}\,{Hz}^{-1/2}}$ reported in~\cite{tu2024approaching}. When we remove all the noise sources in the simulation, the estimated Rx SNR rises to 18.22\,dB, which implies that the noise caused by systematic errors (symbol synchronization error, baseband pulse shaping error, atomic nonlinearity, etc.) in this simulation is larger than the BBR-induced noise plus the electronic thermal noise. If the quantum conductance $g_q$ could be further improved by about 20\,dB due to future experimental advancements, the photocurrent thermal noise is anticipated to be greatly suppressed. In this case, the BBR noise is expected to dominate the total system noise, where the system performance can finally achieve the bound~\eqref{eqn:quantum_receiver_SNR}. 


\subsection{Discrete-time baseband simulation for MIMO communications}
In this subsection, we present the equivalent baseband model of the Rydberg superheterodyne receiver illustrated in Fig.~\ref{fig:sp_circuits}. This baseband model can be directly extended to the MIMO case. By integrating all the dynamic signal models of each processing stage, the continuous-time equivalent baseband model is derived to be  
\begin{equation}
    y_{\rm BB}(t) = \sqrt{\frac{P_{\rm T}}{P_{\rm qref}}} H x_{\rm BB}(t) + w_{\rm BB}(t), 
    \label{eqn:ctime_equiv_bbmodel}
\end{equation}
where $x_{\rm BB}(t)$ is the Tx complex baseband signal of unit power that lies within the 3\,dB bandwidth of $g_q(\ri\omega)$, $y_{\rm BB}(t)$ is the Rx complex baseband signal, $H$ is the channel coefficient in the usual sense\footnote{The channel coefficient is canonically defined as the power-based transfer function $s_{21}$ between the Tx isotropic antenna port and the Rx isotropic antenna port, evaluated at any specified frequency point within the communication bandwidth. }, $P_{\rm T}\,{\rm [W]}$ is the transmitted power of the classical transmitter, $P_{\rm qref}\,{\rm [W]}$ is the quantum reference power level defined by 
\begin{equation}
    \frac{1}{\sqrt{P_{\rm qref}}} = \frac{1}{2V_{\rm ref}}R_{\rm T}K_c L|g_q(\ri 2\pi f_{\rm IF})|\sqrt{\frac{8\pi\eta_0}{\lambda_{c}^2}},  
\end{equation}
and $w_{\rm BB}(t)$ is a complex white process within the instantaneous bandwidth upper-bounded by the bandwidth of $g_q(\ri\omega)$. The noise PSD is given by 
\begin{equation}
    {\rm PSD}[w_{\rm BB}(t)] = n_{\rm w, bbr} + n_{\rm w, shot}+n_{\rm w, TIA}+n_{\rm w, th},  
    \label{eqn:equiv_baseband_noisePSD}
\end{equation}
where the components are computed to be  
\begin{equation}
    n_{\rm w, bbr} = \frac{1}{2}\left(\frac{1}{\sqrt{2}V_{\rm ref}} (R_{\rm T}K_c) Lg_q \sqrt{\zeta(\ell)}\cdot\sqrt{2}E_{\rm n,bb}\right)^2, 
\end{equation}
\begin{equation}
    n_{\rm w, shot} = \frac{1}{2}\left(\frac{1}{V_{\rm ref}} (R_{\rm T}K_c) \sqrt{2q_e \bar{I}_{\rm ph}}\right)^2, 
\end{equation}
\begin{equation}
    n_{\rm w,TIA} = \frac{1}{2}\left(\frac{V_{\rm n, TIA}R_{\rm T}}{ V_{\rm ref}(Z_{\rm in}+R_s)}\right)^2+\frac{1}{2}\left(\frac{I_{\rm n, TIA}K_cR_{\rm T}}{ V_{\rm ref}}\right)^2, 
\end{equation}
\begin{equation}
    n_{\rm w,th} = \frac{1}{2}\left(\frac{K_cR_{\rm T}}{V_{\rm ref}}\sqrt{\frac{4k_{\rm B}T}{R_{s}}}\right)^2. 
\end{equation}
The voltage $V_{\rm ref}$ is the reference voltage of the I/Q baseband ADC. We further assume a communication bandwidth $W$ with the Tx baseband signal represented by $x_{\rm BB}(t) = \sum_n x(n)\sinc(Wt-n)$, then the continuous-time equivalent baseband model~\eqref{eqn:ctime_equiv_bbmodel} is discretized into 
\begin{equation}
    y(n) = \sqrt{P_{\rm T}/P_{\rm qref}} Hx(n) + w(n),
    \label{eqn:dtime_equiv_bbmodel}
\end{equation}
where $y(n)=y_{\rm BB}(n/W)$, and $w(n)$ is an i.i.d. complex white Gaussian sequence of variance $\sigma_w^2=W\cdot{\rm PSD}[w_{\rm BB}(t)]$ computed from~\eqref{eqn:equiv_baseband_noisePSD}. Note that the discrete-time signal model~\eqref{eqn:dtime_equiv_bbmodel} can be easily extended to a MIMO signal model ${\bf y} = \sqrt{P_{\rm T}/P_{\rm qref}} {\bf H}{\bf x}+{\bf w}$ for each single MIMO channel use, or extended to a linear discrete-time multipath channel model $y(n) = \sqrt{P_{\rm T}/P_{\rm qref}} \sum_m H(m)x(n-m)+w(n)$, where $H(m)$ is the channel coefficient evaluated at the $m$-th tap.  

\begin{figure}
    \centering
    \captionred
    \includegraphics[width=\figsize\linewidth]{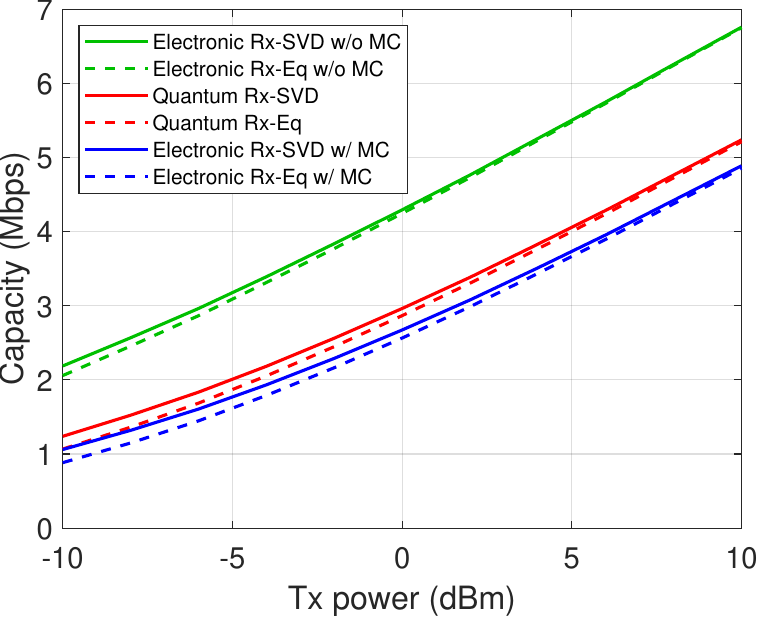}
    \caption{MIMO capacity simulations for the quantum and classical receivers. ``SVD'' denotes water-filling power allocation across MIMO sub-channels with SVD-based Tx precoding/Rx combining, and ``Eq'' denotes equal power allocation.} 
    \label{fig:q-mimo}
\end{figure}

\red{Figure~\ref{fig:q-mimo}} shows the simulated $8\times 8$ MIMO capacity for classical/quantum receivers with instantaneous bandwidth $B=100\,{\rm kHz}$ and Rayleigh channel fading. The noise factor of classical electronic receiver is set to be $F_c=2\,{\rm dB}$. All the other simulation parameters are aligned to the previous continuous-time simulations. For the classical MIMO system, the Rx antenna mutual coupling (MC) effect is considered~\cite{wallace2002capacity}, resulting in a reduction of the received SNR and MIMO degrees-of-freedom (DoF). In contrast, the Rydberg atomic MIMO receiver is not susceptible to the antenna MC effect~\cite{yuan2025electromagnetic}. Thus, the quantum receiver outperforms the MC-influenced classical receiver. 

\section{Conclusions} \label{sec6}
In this paper, we provided a novel LTI system viewpoint for Rydberg atomic receivers by solving the Laplace domain transfer function directly from the physics-rooted master equation. With this LTI system viewpoint as well as the derived transfer functions, the atomic response to the signal field and the noise field can be readily computed in closed-form, which facilitates future study in Rydberg quantum communication and sensing systems. 
Finally, based on the signal and noise models, a single-carrier SISO/MIMO simulator for wireless communications was proposed, where the simulation results demonstrated the possibility of constructing a quantum MIMO receiver that outperforms classical electronic receivers. 

Future works will be focused on experimental validation of the general signal model, and the integration of zero-IF architectures to further mitigate in-band BBR noise, bringing Rydberg atomic receivers closer to their theoretical performance limits. 

\section{Acknowledgment}
The authors would like to thank Prof. Wei E. I. Sha from Zhejiang University, Dr. Hanfeng Wang from Massachusetts Institute of Technology, and Mingyao Cui from The University of Hong Kong for their helpful discussions and constructive suggestions on Rydberg atomic receivers.  

\appendices 
\section{Proof sketch of {\bf Theorem~\ref{theorem1}}} \label{appendix_1}
\red{
To prove this theorem, we need to evaluate the Gaussian-weighted integral~\eqref{eqn:Tkl_with_Doppler} with integration variable $v$. Since the variable $v$ appears in two places: $(v/\sigma_v){\bf C}_v$ and $\bar{\bf z}_v$, we can separate $(v/\sigma_v)$ from the matrix expression by the eigenvalue decomposition (EVD), which is given by  
\begin{equation}
\begin{aligned}
    (s{\bf I}-{\bf C}_0)^{-1}{\bf C}_v &= {\bf S}(s){\rm diag}({\lambda}_1(s), \cdots, {\lambda}_n(s)){\bf S}^{-1}(s) \\
    &= \sum_{m=1}^{15} \lambda_m(s) [{\bf S}(s)]_{:,m}[{\bf S}^{-1}(s)]_{m,:}. 
\end{aligned}
\end{equation}
This EVD simplifies the matrix inversion in~\eqref{eqn:Tkl_s_v} as 
\begin{equation}
\begin{aligned}
    (s{\bf I}_{15}-{\bf C}_0 - (v/\sigma_v){\bf C}_v)^{-1} &= \left({\bf I}-\frac{v}{\sigma_v}(s{\bf I}-{\bf C}_0)^{-1}{\bf C}_v \right)^{-1} \\
    &~~~\times (s{\bf I}-{\bf C}_0)^{-1},
\end{aligned}
\end{equation}
where the $v$-dependent part $\left({\bf I}-\frac{v}{\sigma_v}(s{\bf I}-{\bf C}_0)^{-1}{\bf C}_v \right)^{-1}$ is expressed in EVD as 
\begin{equation}
\begin{aligned}
    &\left({\bf I}-\frac{v}{\sigma_v}(s{\bf I}-{\bf C}_0)^{-1}{\bf C}_v \right)^{-1}  = {\bf S}(s)\times \\ 
    &{\rm diag}\left(\frac{1}{1-{\lambda}_1(s)v/\sigma_v}, \cdots, \frac{1}{1-{\lambda}_n(s)v/\sigma_v}\right){\bf S}^{-1}(s).
\end{aligned}
\end{equation}
Similarly, the $v$-dependent steady state $\bar{\bf z}_v$ is expressed as 
\begin{equation}
\begin{aligned}
    & \bar{\bf z}_v = {\bf S}(0)\times \\ 
    & ~~~{\rm diag}\left(\frac{1}{1-\lambda_1(0)v/\sigma_v}, \cdots, \frac{1}{1-\lambda_{15}(0)v/\sigma_v}\right) {\bf S}^{-1}(0)\bar{\bf z}_0.
\end{aligned}
\end{equation}
We can see that the integrand of $T_{k\ell}^{\rm D}(s)$ depends on a dual-pole function of $v$. This kind of integral is closely related to the following integral $J(z)$ defined as 
\begin{equation}
    J(z) = \int_{-\infty}^\infty \frac{1}{z-\xi}\mathcal{N}(\xi|0,1){\rm d}\xi,\quad z\in\mathbb{C},
\end{equation}
which is a simple pole convolved with a standard Gaussian. For $z\in\mathbb{R}$, the integral $J(z)$ can be evaluated by the definition of the Cauchy principal value integral. When extended to the complex plane, additional residue at $\Re{z}$ should be considered, resulting in the ${\rm sgn}\,\Im{z}$ term in~\eqref{eqn:J(z)_analytical_expression}. The final expression of $\mathbb{E}_X[1/(1-\lambda_m(s)X)(1-\lambda_n(0)X)]$ is obtained by decomposing the dual-pole function of $X=v/\sigma_v$ into the linear combination of two simple poles.  
}

\footnotesize

\bibliographystyle{IEEEtran}
\bibliography{IEEEabrv, bibfile}

\end{document}